\documentclass[aps,prb,twocolumn,groupedaddress,superscriptaddress,showpacs]{revtex4-1}
\usepackage{graphicx}
\usepackage{amsmath}

\begin{document}

\title{Sound behavior near the Lifshitz point in proper ferroelectrics}

\author{A.~Kohutych}
\author{R.~Yevych}
\author{S.~Perechinskii}
\affiliation{Institute for Solid State Physics and Chemistry, Uzhgorod University, Pidgirna Str. 46, Uzhgorod, 88000, Ukraine}
\author{V.~Samulionis}
\author{J.~Banys}
\affiliation{Faculty of Physics, Vilnius University, Sauletekio 9, 10222 Vilnius, Lithuania}
\author{Yu.~Vysochanskii}
\email{vysochanskii@univ.uzhgorod.ua}
\affiliation{Institute for Solid State Physics and Chemistry, Uzhgorod University, Pidgirna Str. 46, Uzhgorod, 88000, Ukraine}

\date{\today}

\begin{abstract}
The interaction between soft optic and acoustic phonons was investigated for Sn$_2$P$_2$(Se$_{0.28}$S$_{0.72}$)$_6$ proper uniaxial ferroelectrics by Brillouin scattering and ultrasonic pulse--echo techniques. The elastic softening of hypersound velocity of transverse acoustic phonons and for both longitudinal and transverse ultrasound waves which propagate near direction of the modulation wave vector (in the incommensurate phase at $x\!>\!x_{LP}$) was found at cooling to the Lifshitz point in the paraelectric phase. The strong increase of the ultrasound attenuation have also been observed. Such phenomena are related to the linear interaction of the soft optic and acoustic branches in the region of relatively short--range hypersound waves and to the strongly developed long--range order parameter fluctuations in the ultrasound frequency range. The hypersound velocity temperature dependence was described within the Landau--Khalatnikov approximation for the ferroelectric phase. \end{abstract}

\maketitle

\section{Introduction}
For mixed Sn$_2$P$_2$(Se$_x$S$_{1-x}$)$_6$ crystals, the Lifshitz point (LP) is presented near $x_{LP}\!\approx\!0.28$ and $T_{LP}\!\approx\!284$~K on the temperature--concentration phase diagram.\cite{bib1_1,*bib1_2} This multicritical point divides the line $T_0(x)$ of the second order phase transitions, between paraelectric (P2$_1$/c) and ferroelectric (Pc) phases at $x\!<\!x_{LP}$, from the line $T_i(x)$ of the second order phase transitions from the paraelectric phase into incommensurate (IC) phase at $x\!>\!x_{LP}$.\cite{bib2_1,*bib2_2} This IC phase is of type II and related to the Lifshitz like invariant in the thermodynamic potential density expansion of the order parameter and its derivatives with an account of elastic energy and interaction between polarization and deformations.\cite{bib3_1,*bib3_2,*bib3_3} At approaching to the LP, both temperature width of IC phase $T_i\!-\!T_c$ and wave vector $q_i$ of the spontaneous polarization modulation along the line $T_i(x)$ continuously decrease to zero: $T_i\!-\!T_c\!\sim\!(x\!-\!x_{LP})^2$ and $q_i\!\sim\!(x\!-\!x_{LP})^{1/2}$.\cite{bib2_1,*bib2_2,bib4_1,*bib4_2} The exponents of critical behavior near the LP with one direction of modulation in uniaxial ferroelectrics could be modified by developed critical fluctuations.\cite{bib5,bib6_1,*bib6_2,bib7} In this case, the pillow-like anisotropy of the critical fluctuations in the reciprocal space was predicted theoretically\cite{bib5} and have been confirmed by synchrotron radiation diffuse scattering experiments.\cite{bib8} Also, two--length--scale critical phenomenon have been predicted\cite{bib8} in uniaxial Sn$_2$P$_2$S$_6$ (SPS) ferroelectrics with second order phase transition near the LP on the state diagram. The observed shape of the phase diagram and the X--ray diffraction data for temperature and concentration dependencies of the modulation wave vector in IC phase of Sn$_2$P$_2$(Se$_x$S$_{1-x}$)$_6$ ferroelectrics generally satisfy predicted behavior for the case of LP.\cite{bib9_1,*bib9_2} From these experiments, the diffraction data about modulation behavior are available only far enough from the $x_{LP}\!\approx\!0.28$, i.e. for $x\!=\!0.6$, 0.8 and 1,\cite{bib2_1,*bib2_2} because the related satellites were not resolved for smaller concentrations of selenium in the Sn$_2$P$_2$(Se$_x$S$_{1-x}$)$_6$ mixed crystals.

The wave vector $q_i$ position near the Brillouin zone center is determined by linear interaction of the soft optic and acoustic phonon branches with similar symmetry. Such interaction was clearly observed by inelastic neutron scattering for the Sn$_2$P$_2$Se$_6$  crystals.\cite{bib10} It was estimated that $q_i$ could move to zero at decreasing $x$ to $x_{LP}$ mostly on the matter of changing in the soft optic branch. It is expected that the optic--acoustic phonon interactions and developed fluctuations could strongly influence on the acoustic properties of investigated crystals in the nearest vicinity of the LP. The anomalies of acoustic properties could be observed by Brillouin scattering spectroscopy and by ultrasonic measurements.

The linear interaction between optic and acoustic phonons was earlier observed by inelastic neutron scattering for cubic paraelectric phases of perovskite crystals SrTiO$_3$, BaTiO$_3$, KNbO$_3$, KTaO$_3$\cite{bib11,bib12,bib13} and for crystalline SiO$_2$.\cite{bib14} Such interaction is supposed to be important for the low temperature anomalous behavior of quantum paraelectrics\cite{bib15_1,*bib15_2} and for the description of relaxational dynamics in relaxor materials.\cite{bib16} It is directly involved into mechanism of the IC phase appearance in quartz.\cite{bib14}

The linear repulsion between optic and acoustic branches was investigated for cubic SrTiO$_3$\cite{bib17} and for hexahonal crystals of quartz near phase transition into their IC phase\cite{bib14} by Brillouin scattering spectroscopy. Such repulsion rapidly decreases as $q^2$ at approaching to Brillouin zone center. By this, at Brillouin scattering with participation of phonons with $q\!\approx\!10^5$~cm$^{-1}$, the linear interaction of optic $\omega_0(q)$ and acoustic $\omega_a(q)$ branches is found as weak anticrossing features for the frequencies wave vector dependencies, but was observed clearly for their damping evolution as function of wave vector $q$.\cite{bib14,bib17}

For ultrasound waves with $q\!\approx\!10^2$~cm$^{-1}$, some effects of linear interaction of unstable optic mode and acoustic phonons obviously will appear in very small, practically very difficult for observation, temperature interval of paraelectric phase near the phase transition point. But near the LP even in uniaxial ferroelectrics, the growing of fluctuation effects is expected\cite{bib5,bib6_1,*bib6_2,bib7}, and here some ultrasound wave softening and rise of their attenuation could be really observed.\cite{bib18,bib19,bib20}

For SPS crystals in the paraelectric phase near the second order ferroelectric transition at $T_0\!\approx\!337$~K, the weak logarithmic fluctuational corrections were observed for longitudinal ultrasound velocity\cite{bib21} which are in agreement with investigations of heat capacity critical behavior\cite{bib22} and with renorm--group theoretical predictions.\cite{bib7} The hypersound velocity and attenuation temperature behavior in ferroelectric phase have been studied by Brillouin spectroscopy for these crystals.\cite{bib23} As for ultrasound frequencies\cite{bib21}, the hypersound velocity temperature dependence at $T\!<\!T_0$ was quantitavely described\cite{bib23} in the Landau--Khalatnikov (L--K) model\cite{bib24} with a single relaxation time and electrostriction coefficients as coupling parameters between elastic waves and spontaneous polarization. For description of the hypersound attenuation temperature and frequency dependencies in the ferroelectric phase of SPS crystals, the mode Gruneisen coefficients for the lowest energy optic phonons and for acoustic phonons were applied as interaction parameters in the L--K model.\cite{bib25_1,*bib25_2}

By Brillouin scattering in different geometries, in comparison with ultrasound data, the dispersion of phase sound velocity was investigated for paraelectric phase of SPS and Sn$_2$P$_2$(Se$_{0.28}$S$_{0.72}$)$_6$ crystals in [001] crystallographic direction, which is oriented near the modulation wave vector of the IC phase at $x\!>\!x_{LP}$. Any velocity dispersion was not found for SPS crystals, but for the mixed crystal with $x$=0.28, the dispersion of longitudinal sound velocity have been observed at room temperature in paraelectric phase near the LP.\cite{bib26} This dispersion could be related to the linear interaction of unstable soft optic mode with acoustic phonons that induce the IC phase at $x\!>\!x_{LP}$.

In this paper, the results of Brillouin back scattering spectroscopy together with ultrasound investigations for the Sn$_2$P$_2$(Se$_{0.28}$S$_{0.72}$)$_6$ crystals, of the $x_{LP}$ concentration, in a wide temperature interval including the LP are presented. The experimental data of temperature dependence of hypersound and ultrasound velocity and attenuation for longitudinal and transverse acoustic waves were analyzed with consideration of linear interaction of the soft optic and acoustic phonon branches and with accounting of fluctuation contribution to explain the temperature anomalies in paraelectric phase. The L--K model was used for the acoustic properties analysis in the ferroelectric phase.

The obtained data on acoustic anomalies support earlier conclusion about LP presence at $x_{LP}\!\approx\!0.28$ in Sn$_2$P$_2$(Se$_x$S$_{1-x}$)$_6$ ferroelectrics where X--ray and neutron diffraction data do not permit exact determination of modulation wave in the IC phase at concentration $x\!\rightarrow\!x_{LP}$.

\section{Experimental results}

Brillouin scattering was investigated in back scattering geometry using a He--Ne laser, and a pressure--scanned three--pass Fabry--Perot interferometer with sharpness of 35 and free spectral range of 2.51~cm$^{-1}$. The samples were placed in a UTREX cryostat in which the temperature was stabilized with an accuracy of 0.3~K. With earlier determined\cite{bib25_1,*bib25_2} apparature function 0.04~cm$^{-1}$ and at fitting of sattelite spectral lines by Lorentzians, the velocity $V$ and attenuation $\alpha$ of acoustic phonons were calculated using relations for back scattering geometry:
\begin{eqnarray}
  V &=& \frac{\lambda_0\Delta\nu_{Br}}{2n} \label{eq1}\\
  \alpha &=& \frac{\Gamma_{Br}\pi}{V} \label{eq2}
\end{eqnarray}
where $\Delta\nu_{Br}$  and $\Gamma_{Br}$ is a Brillouin component shift and halfwidth, respectively, $\lambda_0$ is the wave length of the He--Ne laser, $n$ is the refractive index. The accuracy was about 3\% for sound velocities and about 10\% for attenuation. 	In calculations of the hypersound velocities, the refractive index $n\!=\!3.25$ for the Sn$_2$P$_2$(Se$_{0.28}$S$_{0.72}$)$_6$ crystals was used.\cite{bib38}

The measurements of the ultrasonic velocity were performed using a computer controlled pulse--echo equipment.\cite{bib27} The precision of relative velocity measurements was better than 10$^{-4}$. The temperature stabilization was better than 0.02~K. The sample was carefully polished to have precisely parallel faces normal to the Z axis. Silicon oil and Nonaq stopcock grease were used as an acoustic bonds for longitudinal and transversal ultrasonic waves respectively. The measurements were carried out at 10~MHz frequency using piezoelectric LiNbO$_3$ transducers.

The investigated Sn$_2$P$_2$(Se$_{0.28}$S$_{0.72}$)$_6$ monocrystals were grown by vapor--transport method.\cite{bib1_1,*bib1_2} For Brillouin scattering, the samples with $10\!\times\!12\!\times\!3$~mm$^3$ dimensions were used. For the ultrasound measurements, the samples with $6\!\times\!4.5\!\times\!4$~mm$^3$ for longitudinal waves and $6\!\times\!4\!\times\!2.3$~mm$^3$ for transversal waves were prepared. Investigated elastic waves were propagated along [001] direction (Z axis) in the monoclinic symmetry plane (010). For studied ferroelectrics, spontaneous polarization vector is directed in the symmetry plane near [100] direction (X axis). The wave vector of modulation in the IC phase (at $x\!>\!x_{LP}$) is oriented in the symmetry plane near [001] direction.\cite{bib2_1,*bib2_2}

\begin{figure}
 \includegraphics*[width=3.4in]{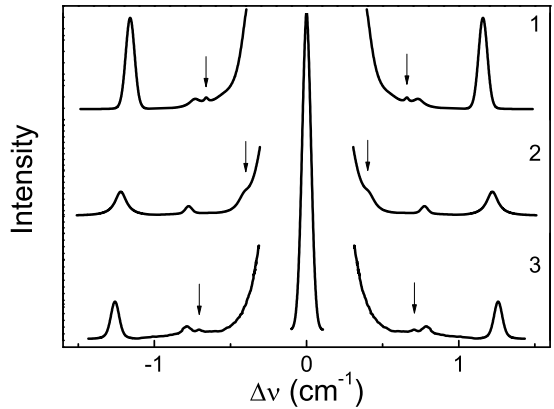}%
 \caption{Brillouin scattering spectra for Sn$_2$P$_2$(Se$_{0.28}$S$_{0.72}$)$_6$ crystal at $Z(X0)\overline{Z}$ geometry for temperatures: 1 -- 203~K, 2 -- 290~K, 3 -- 323~K. The satellites of TA(ZX) phonons are indicated by arrows.\label{fig1}}
\end{figure}

\begin{figure}
 \includegraphics*[width=3.4in]{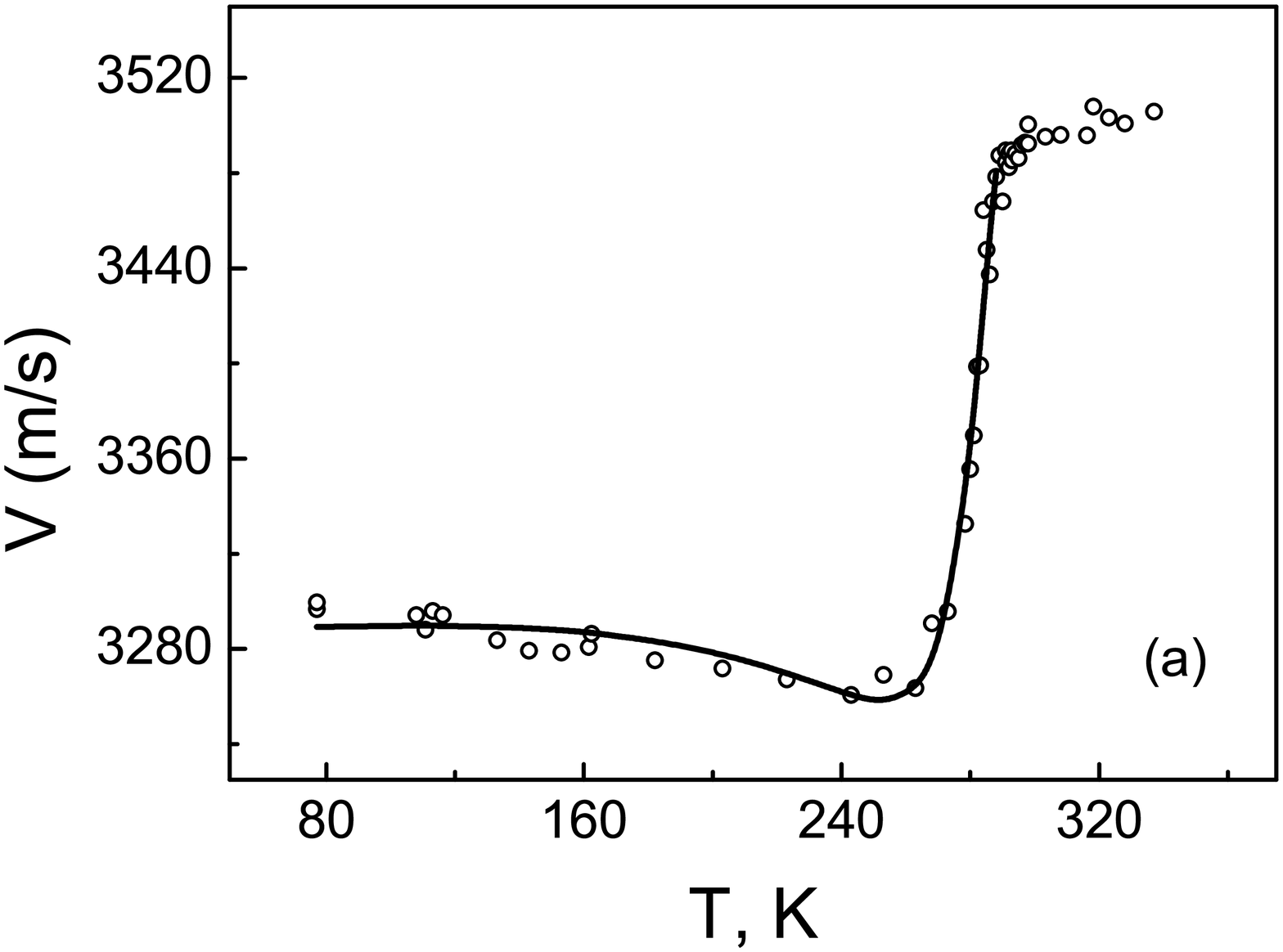}
 \includegraphics*[width=3.4in]{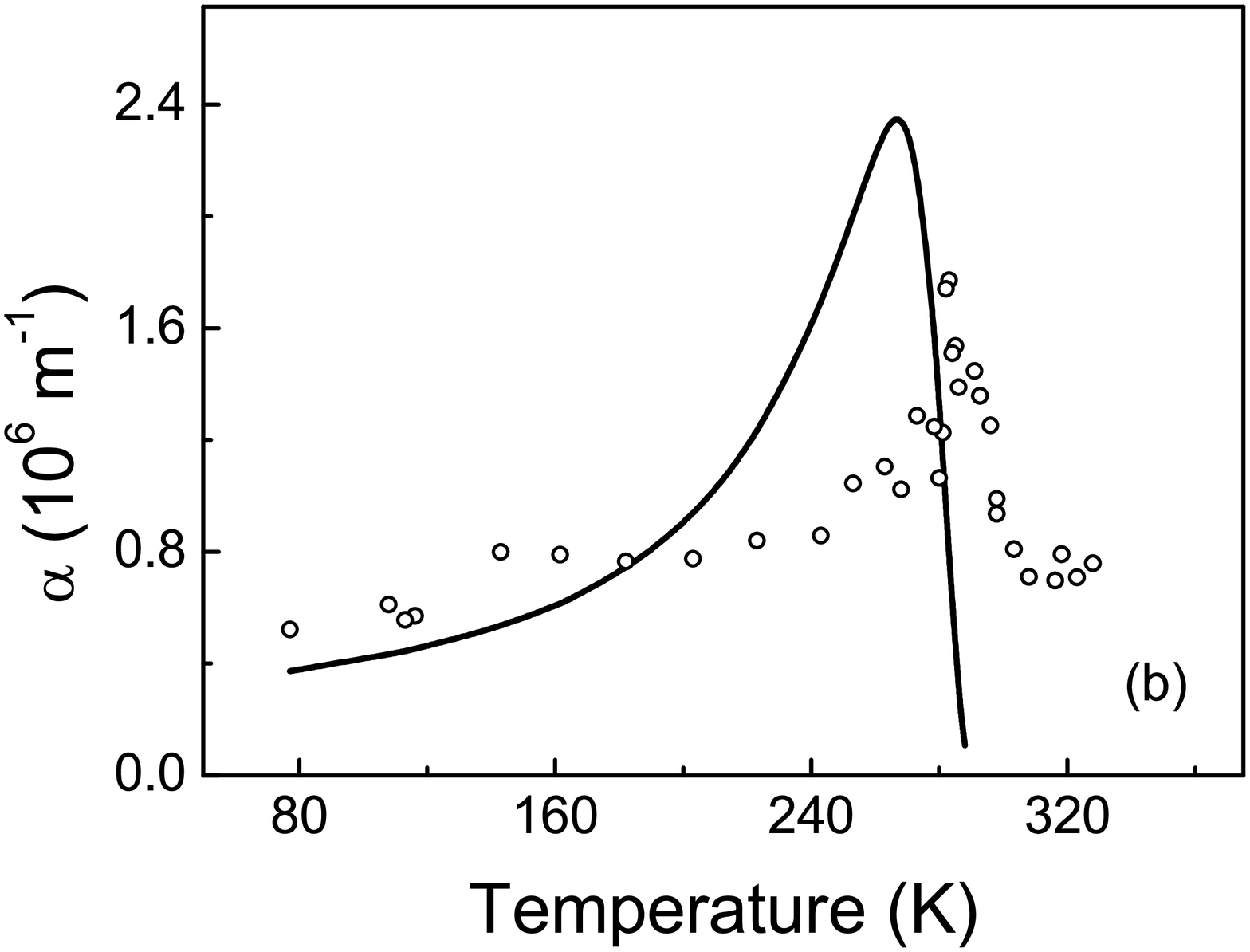}%
 \caption{Temperature dependencies of hypersound velocity (a) and attenuation (b) for longitudinal LA(ZZ) phonons in Sn$_2$P$_2$(Se$_{0.28}$S$_{0.72}$)$_6$ crystal: point --- Brillouin scattering data, solid lines --- calculated dependencies by relations (\ref{eq10}) and (\ref{eq11}) in L--K model.\label{fig2}}
\end{figure}

\begin{figure}
 \includegraphics*[width=3.4in]{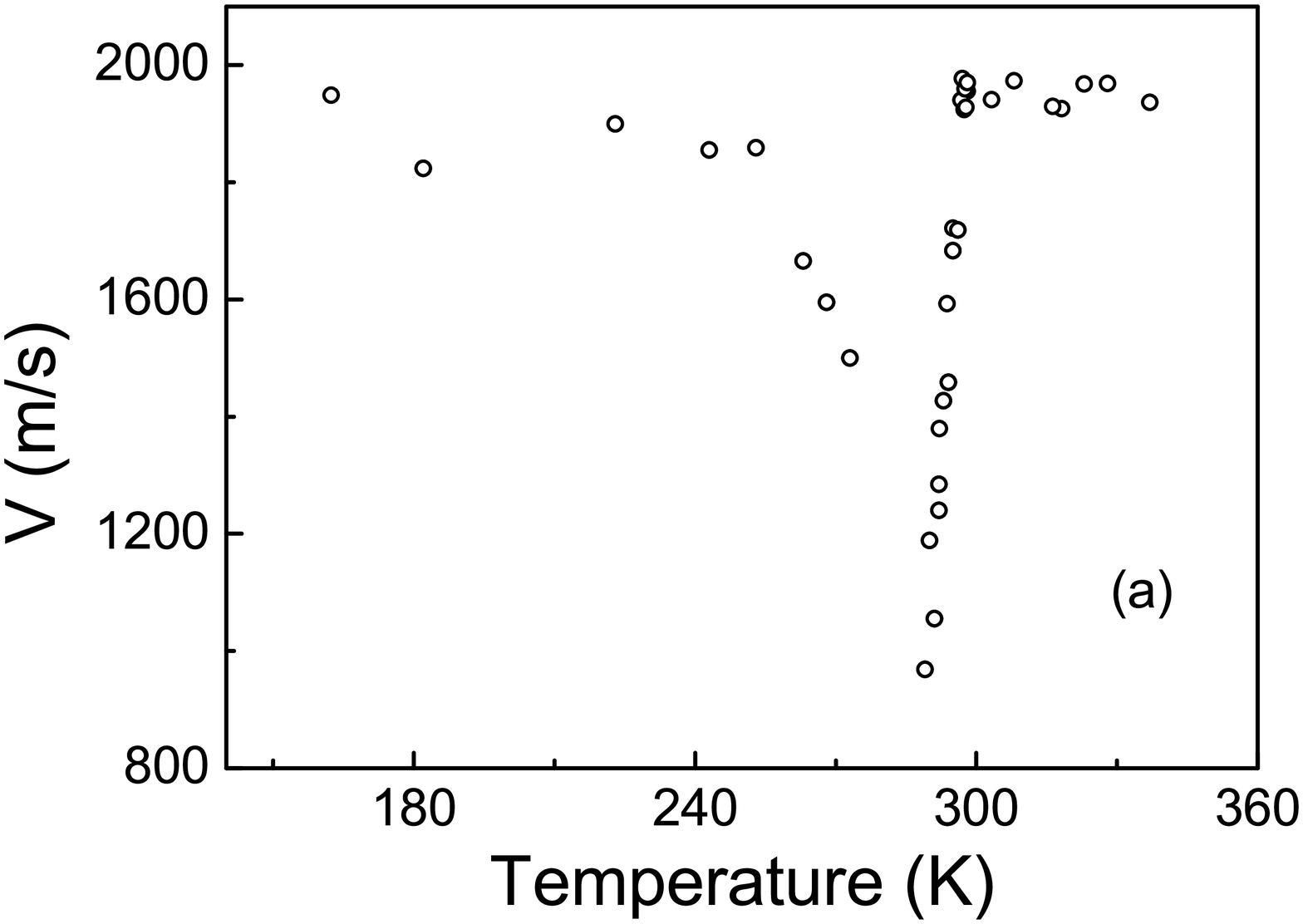}
 \includegraphics*[width=3.4in]{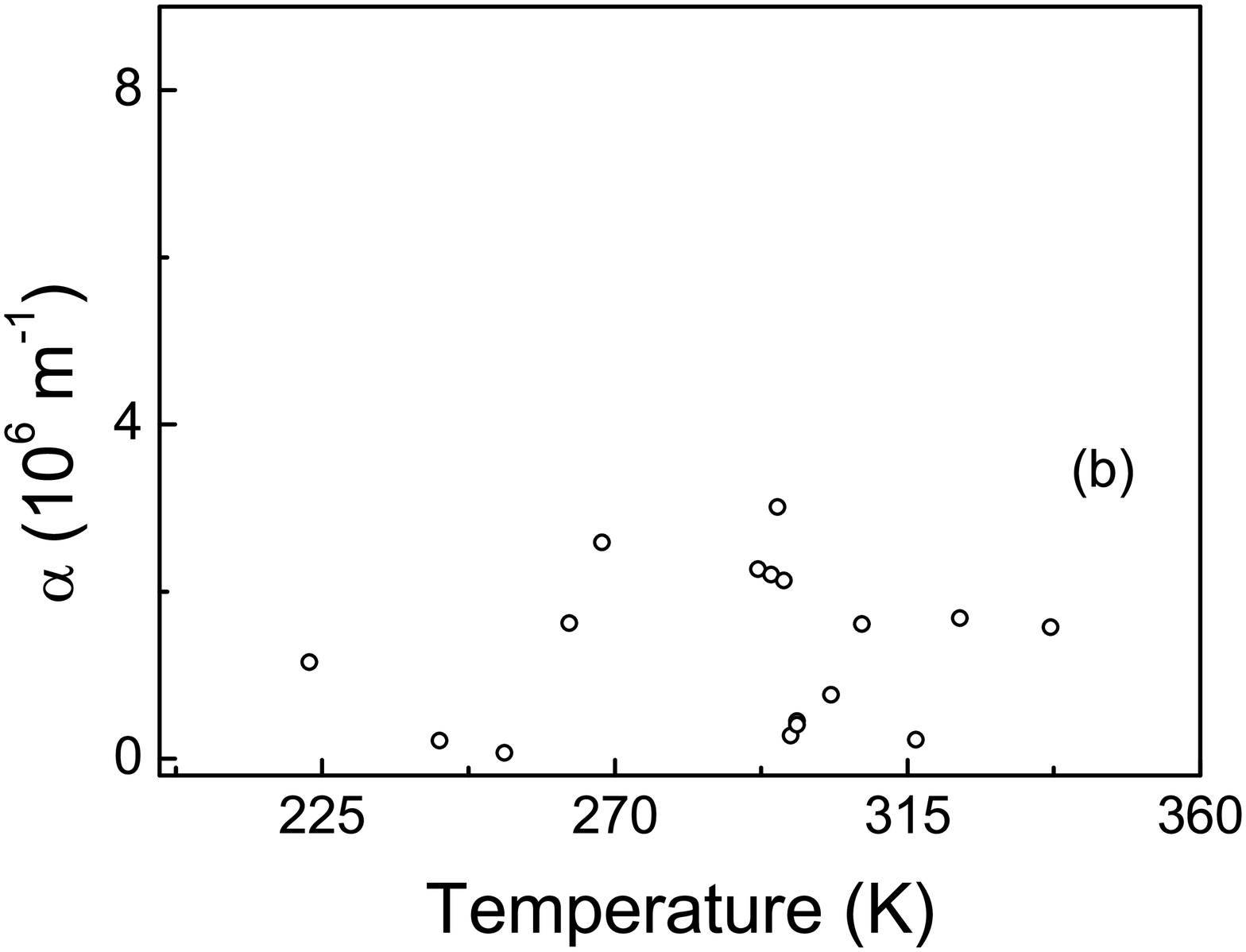}%
 \caption{Temperature dependencies of hypersound velocity (a) and attenuation (b) for transverse TA(ZX) phonons in Sn$_2$P$_2$(Se$_{0.28}$S$_{0.72}$)$_6$ crystal on Brillouin scattering data.\label{fig3}}
\end{figure}

Recorded in $Z(X0)\overline{Z}$ geometry Brillouin scattering spectra for Sn$_2$P$_2$(Se$_{0.28}$S$_{0.72}$)$_6$ crystal at several temperatures are presented at Fig.\ref{fig1}. Related to the longitudinal LA and transverse TA acoustic phonons temperature changes of satellites is clearly seen. The temperature dependencies of phase velocity $V(T)$ and attenuation $\alpha(T)$ for longitudinal LA(ZZ) and transverse TA(ZX) hypersound waves show major anomalies near second order phase transition at $T_0\!\approx\!284$~K (Fig.\ref{fig2} and Fig.\ref{fig3}).

\begin{figure}
 \includegraphics*[width=3.4in]{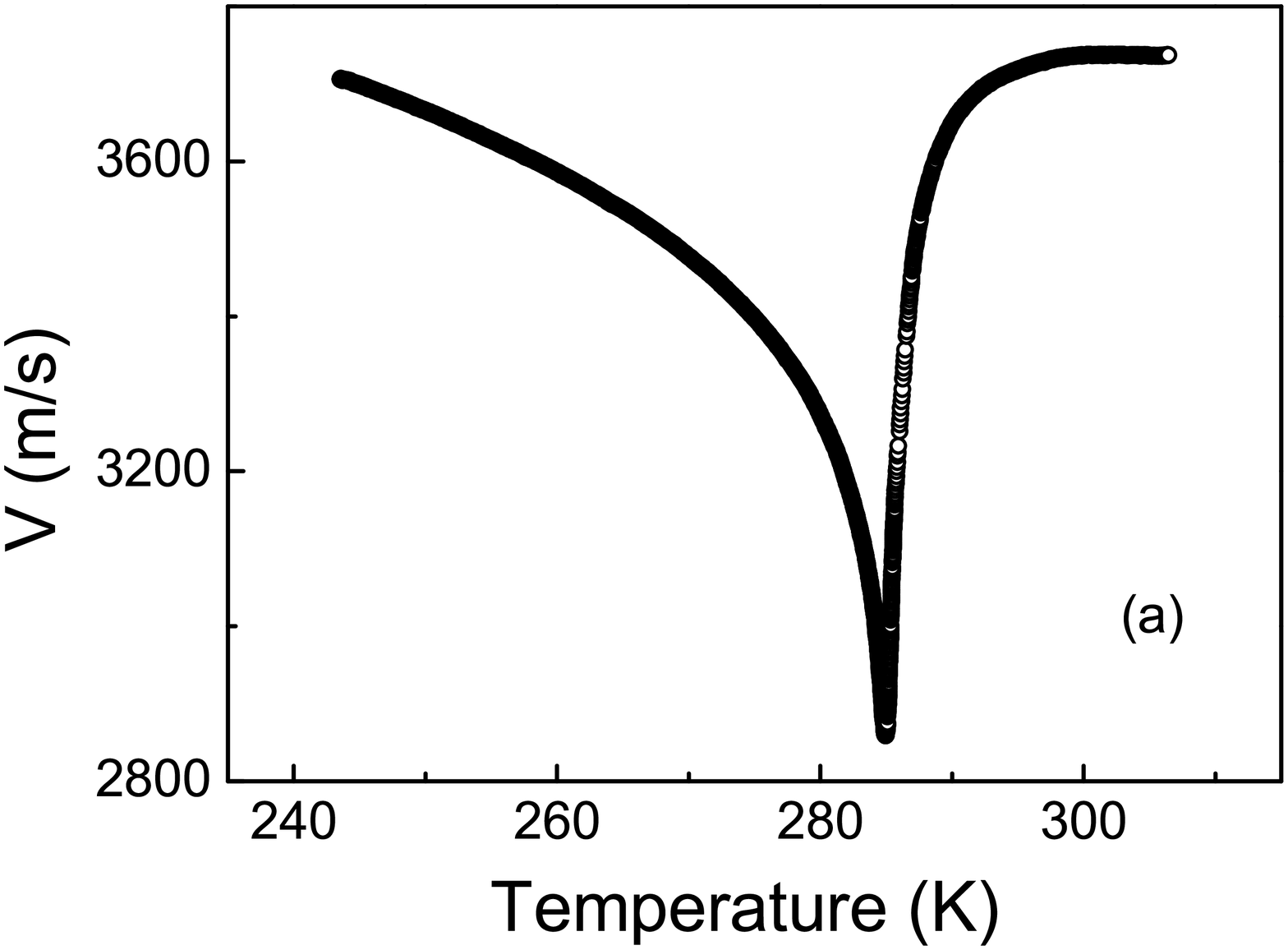}
 \includegraphics*[width=3.4in]{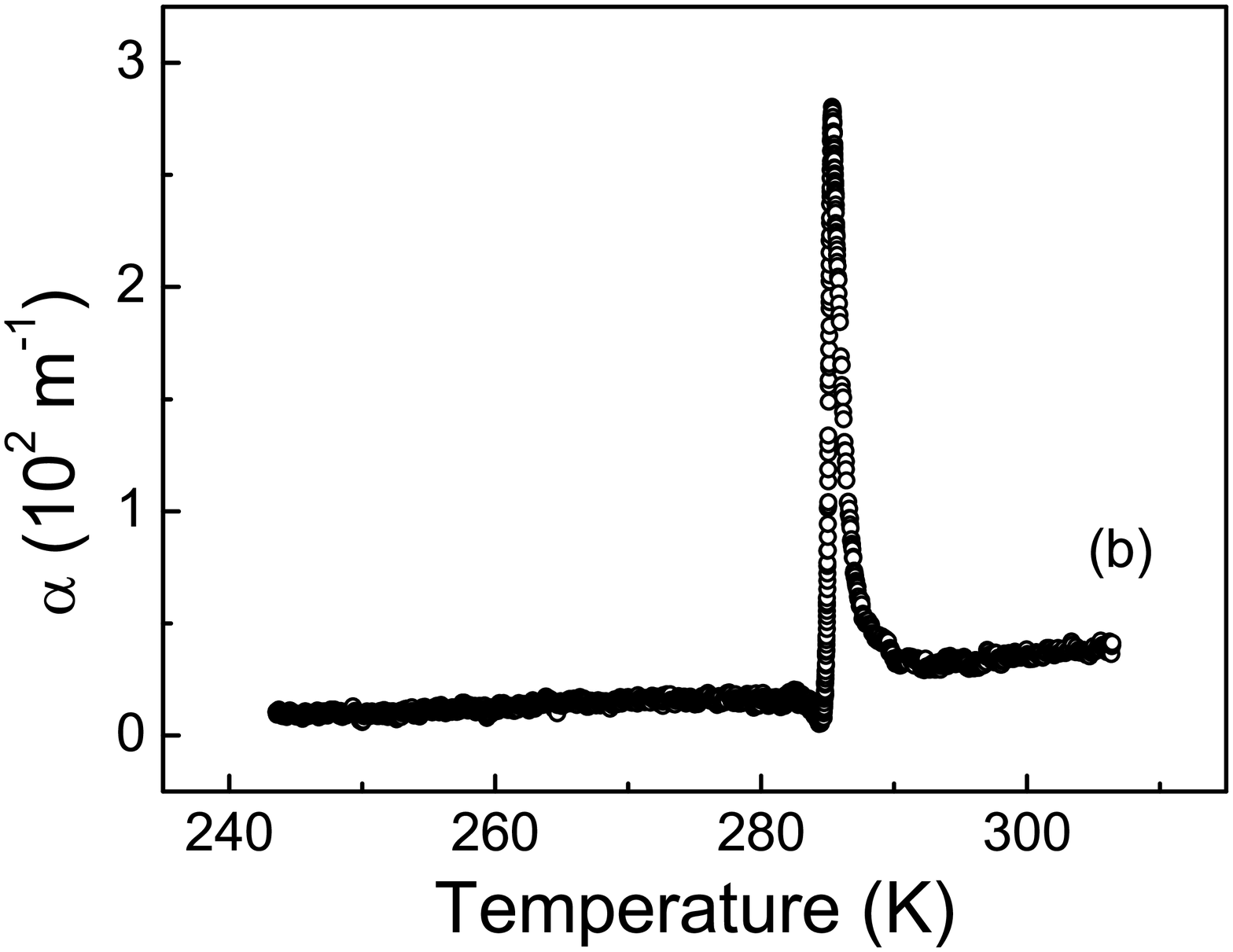}%
 \caption{The velocity (a) and attenuation (b) of longitudinal ultrasonic wave along Z axis temperature dependencies in Sn$_2$P$_2$(Se$_{0.28}$S$_{0.72}$)$_6$ crystal near the LP.\label{fig4}}
\end{figure}

\begin{figure}
 \includegraphics*[width=3.4in]{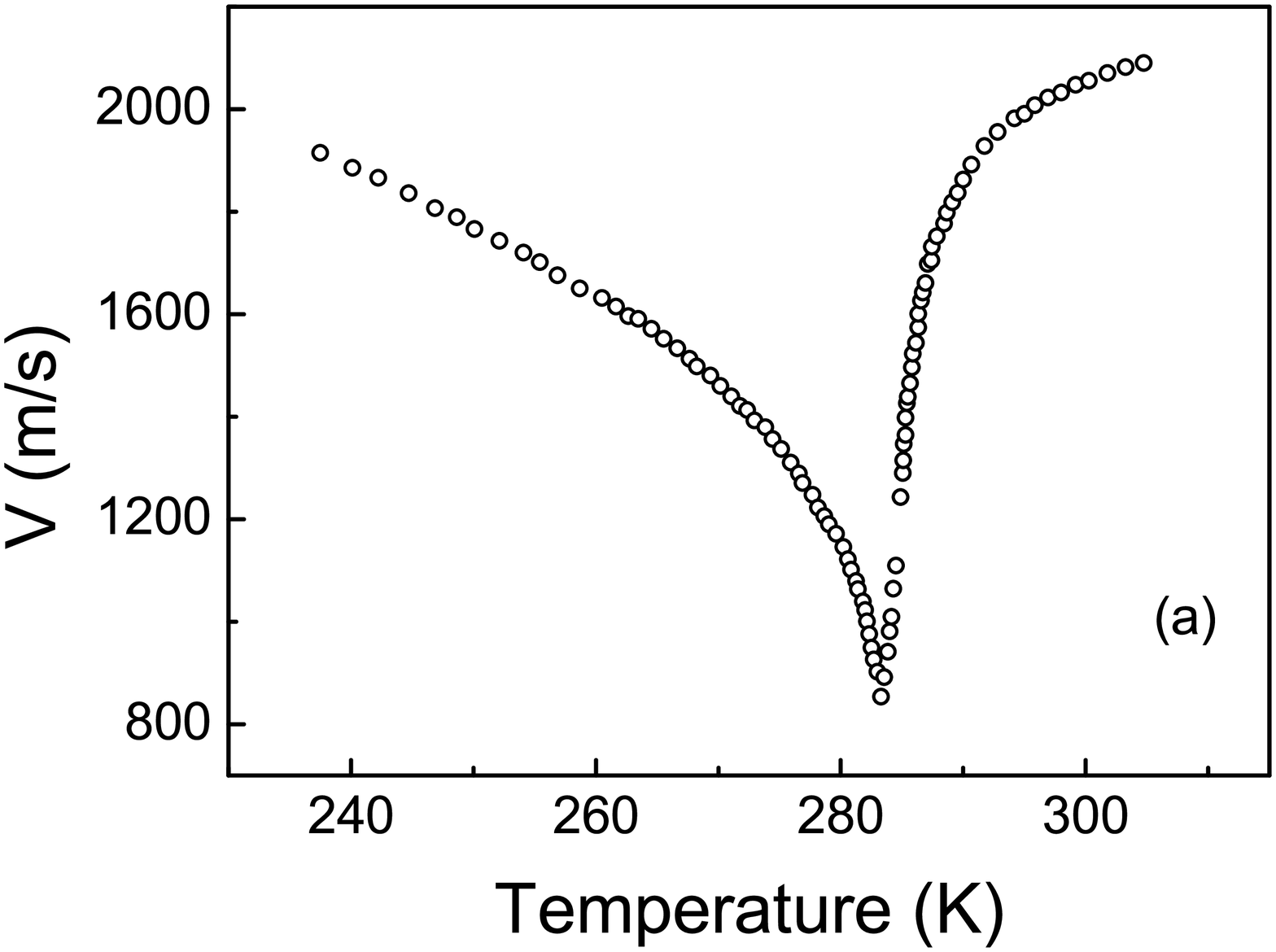}
 \includegraphics*[width=3.4in]{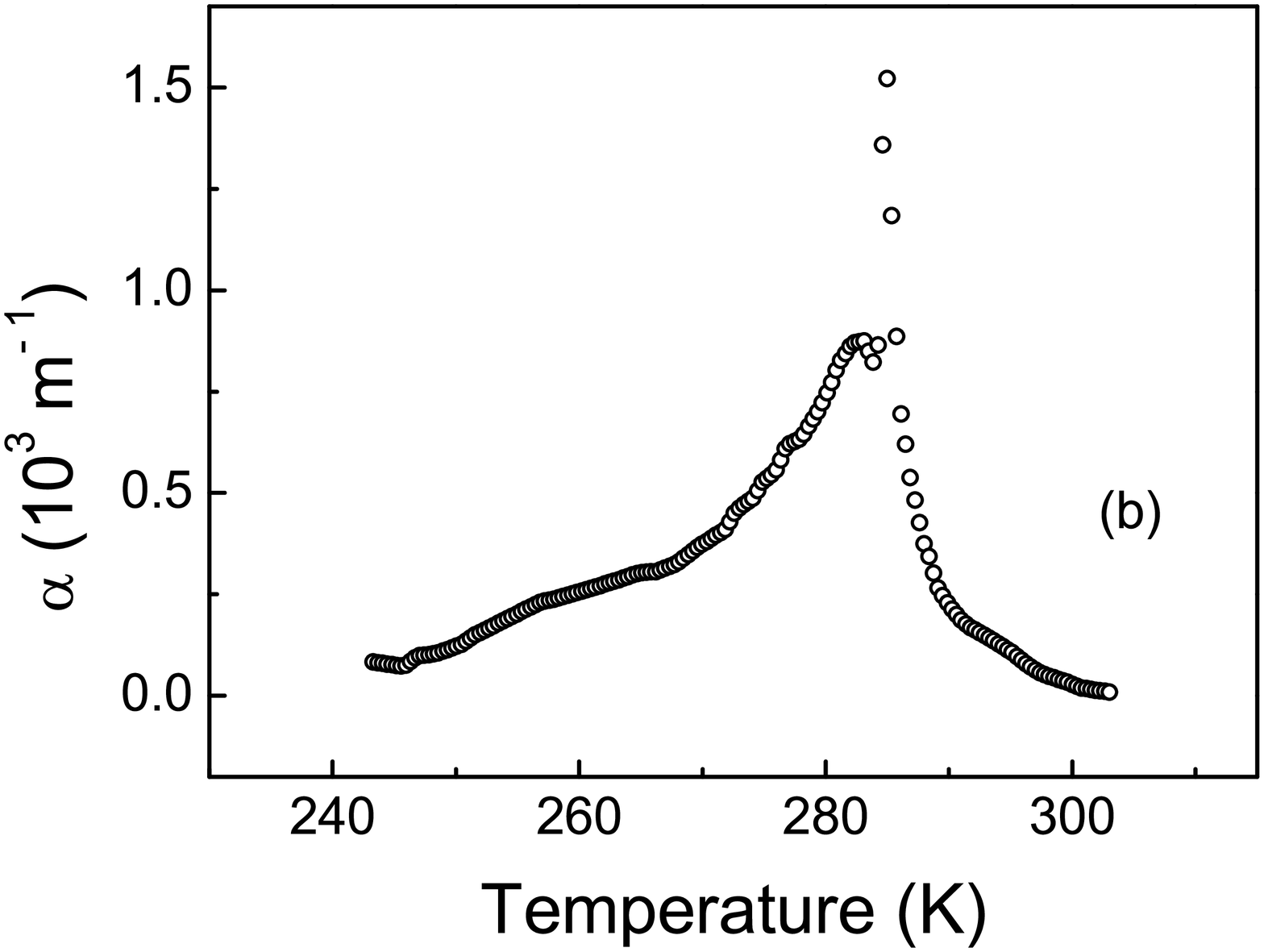}%
 \caption{The temperature dependencies of shear ZX ultrasonic mode (Z--propagation direction, X--displacement) velocity (a) and attenuation (b) near the LP in Sn$_2$P$_2$(Se$_{0.28}$S$_{0.72}$)$_6$ crystal.\label{fig5}}
\end{figure}

The ultrasound investigation data for Sn$_2$P$_2$(Se$_{0.28}$S$_{0.72}$)$_6$ crystals also shows pronounced anomalies for the velocity and attenuation of longitudinal (Fig.\ref{fig4}) and transverse (Fig.\ref{fig5}) acoustic waves. The attenuation anomalies at $T_0$ is so sharp that apparatus can not follow it. So, the attenuation is rounded and at maximum can be 10 or even 30~cm$^{-1}$, but at temperature scale less than 1~mK. Also, the phase transition temperature can be spread over volume of sample because of imperfections or even temperature stability. It is very difficult task to measure such large changes in mKelvins range. Moreover, the attenuation behavior obviously is affected by domain scattering of acoustic waves in the ferroelectric phase.

\begin{figure}
 \includegraphics*[width=3.4in]{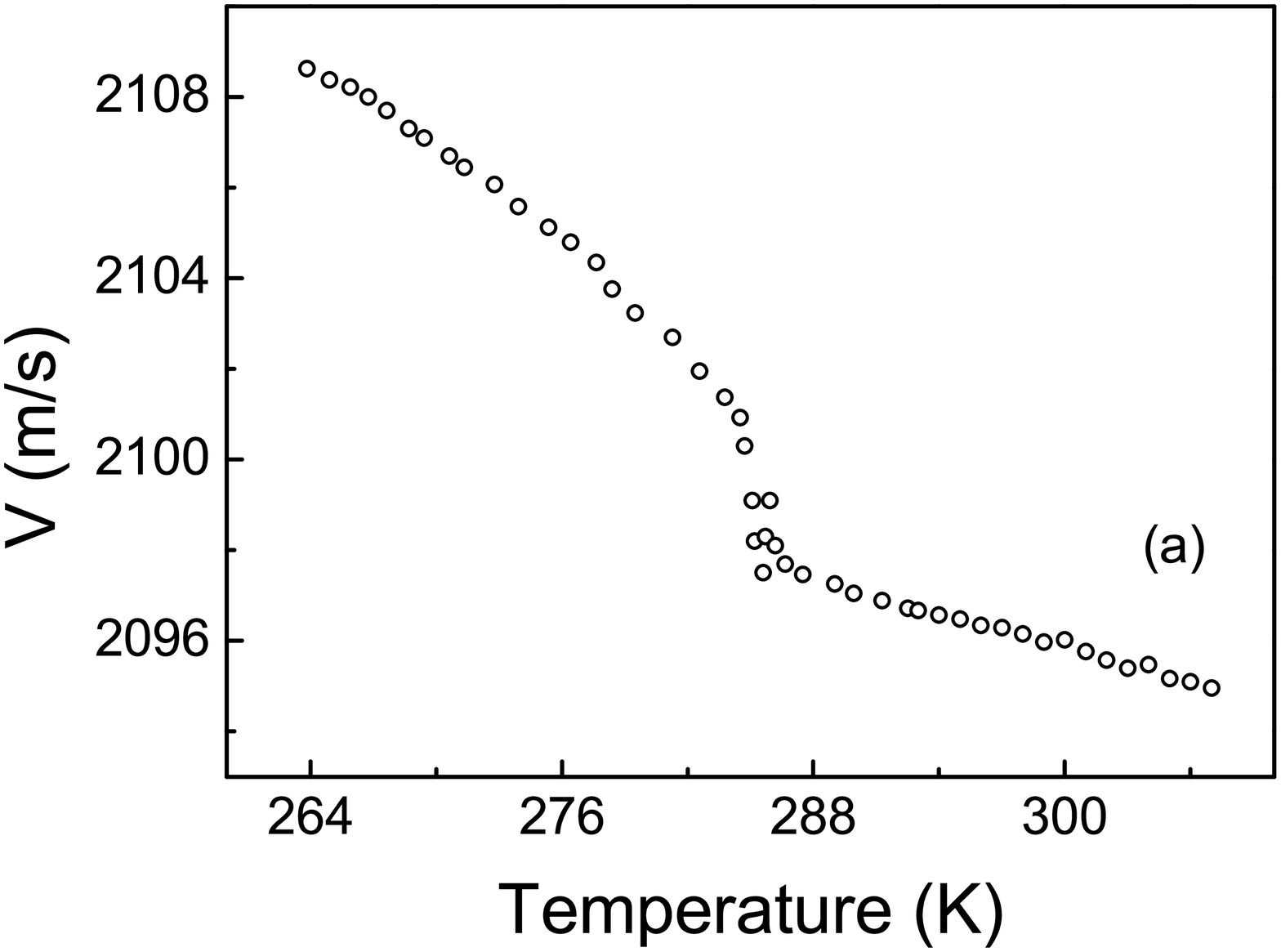}
 \includegraphics*[width=3.4in]{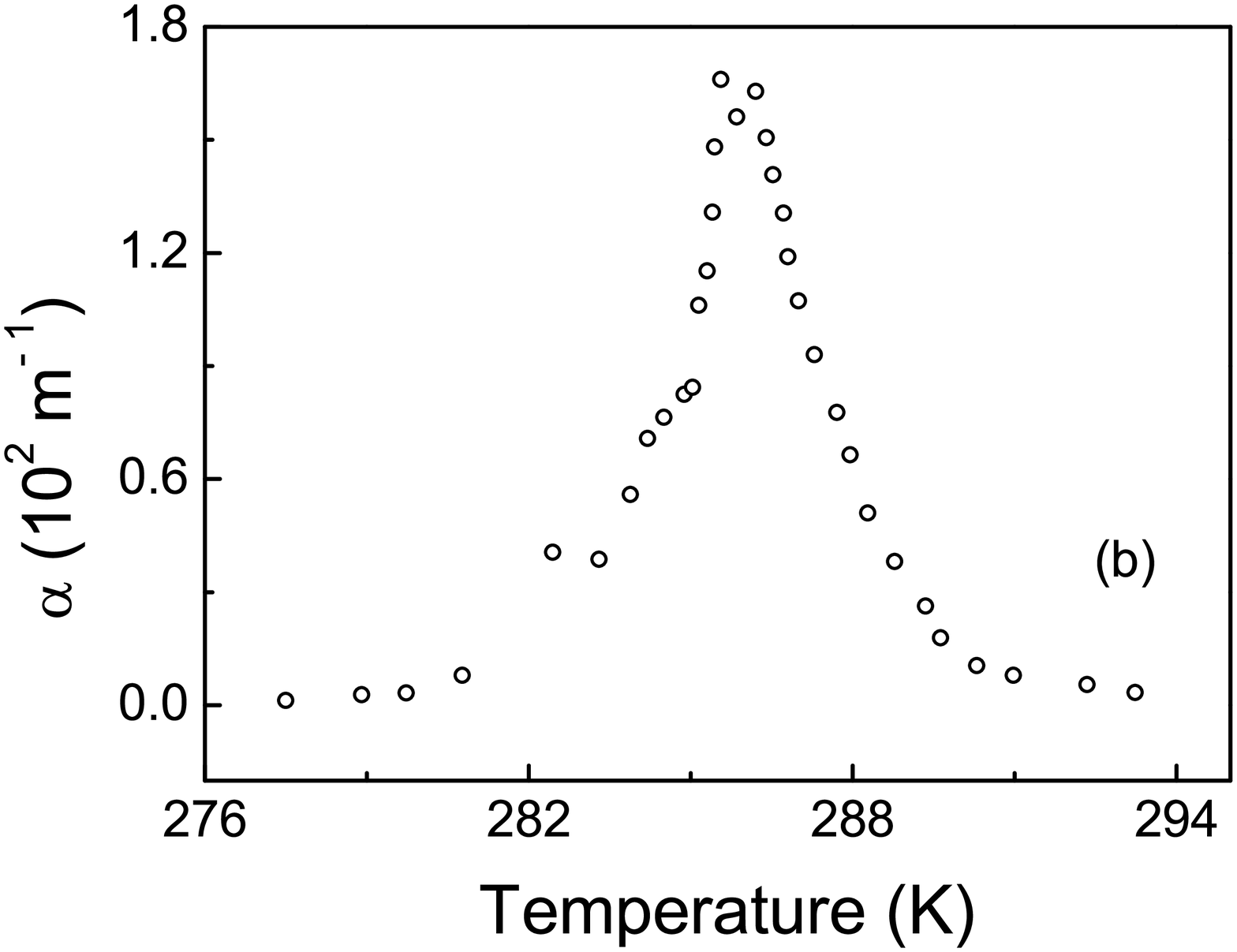}%
 \caption{The temperature dependencies of shear ZY ultrasonic mode (Z--propagation direction, Y--displacement) velocity (a) and attenuation (b) Sn$_2$P$_2$(Se$_{0.28}$S$_{0.72}$)$_6$ crystal.\label{fig6}}
\end{figure}

The ultrasound velocity and attenuation temperature anomalies are largest for the shear TA(ZX) waves (Fig.\ref{fig5}). Contrary, the shear TA(ZY) ultrasound waves demonstrate only small jump for their velocity temperature dependence and a little rise of attenuation near the phase transition temperature (Fig.\ref{fig6}).

\begin{figure}
 \includegraphics*[width=3.4in]{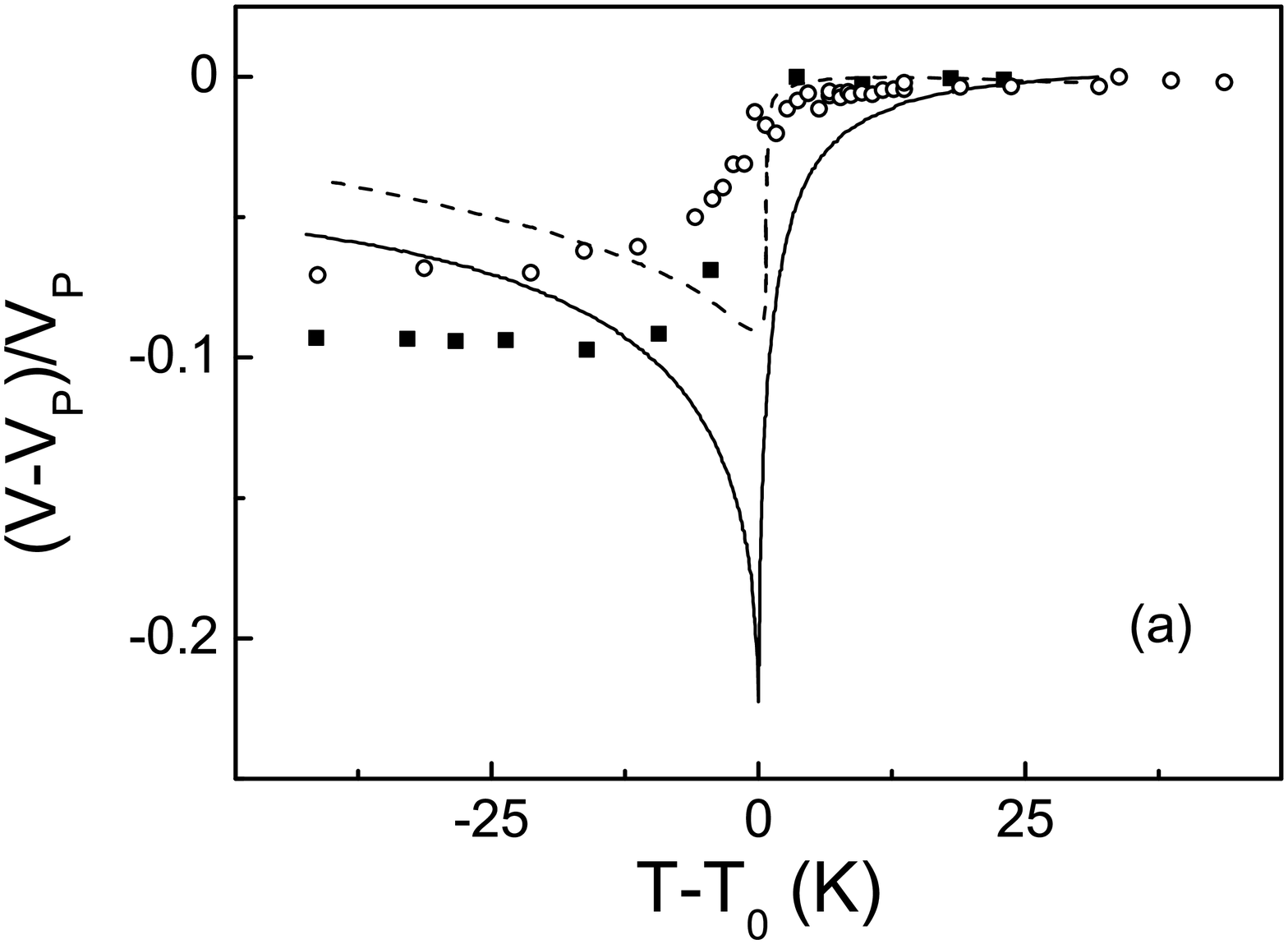}
 \includegraphics*[width=3.4in]{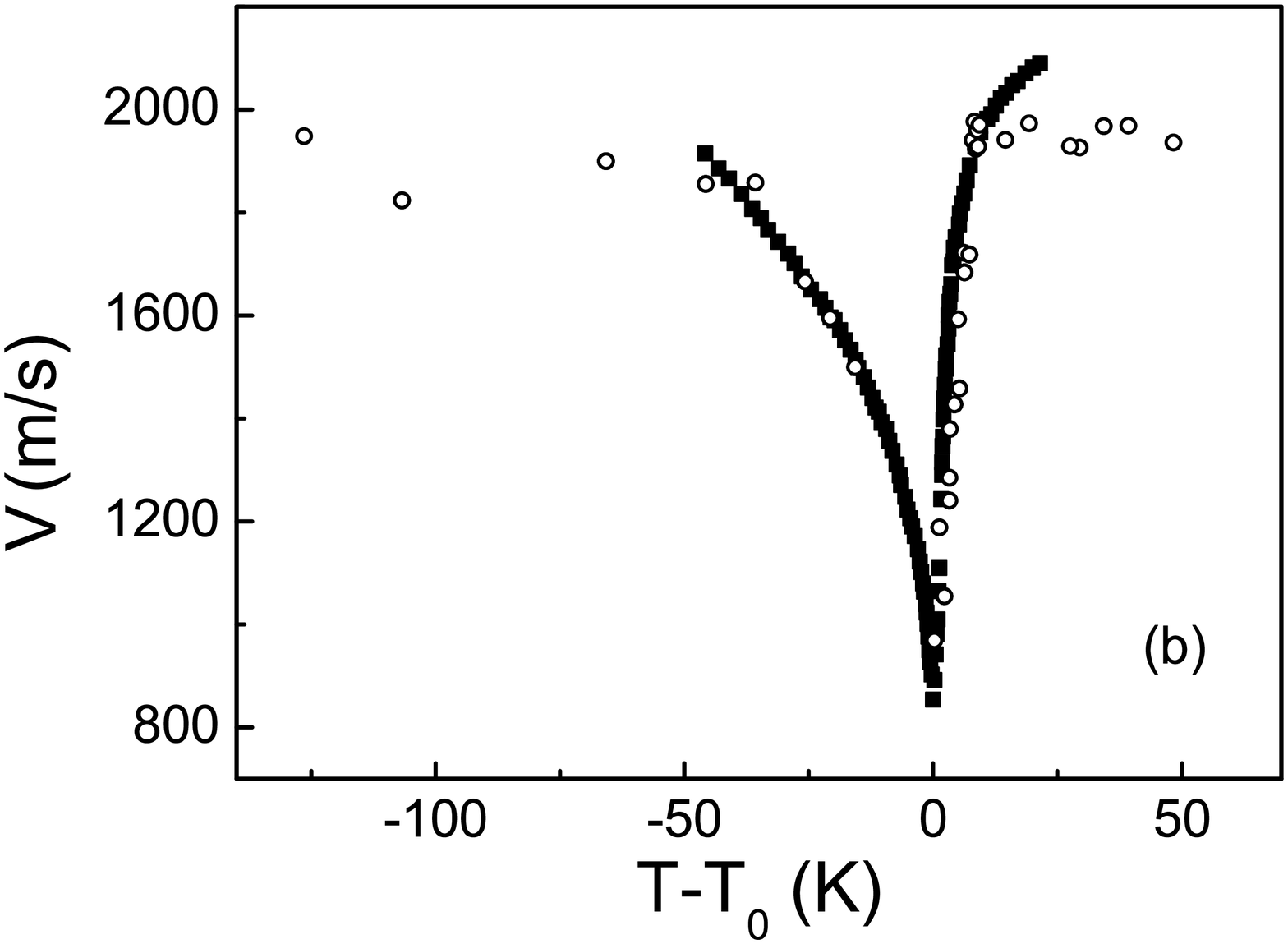}%
 \caption{a -- temperature dependence near the phase transitions of longitudinal LA(ZZ) sound velocity variation relatively their high temperature value $V_p$  in paraelectric phase: for ultrasound in Sn$_2$P$_2$(Se$_{0.28}$S$_{0.72}$)$_6$ (solid line) and in Sn$_2$P$_2$S$_6$ (dashed line); for hypersound in Sn$_2$P$_2$(Se$_{0.28}$S$_{0.72}$)$_6$ (open circles) and in Sn$_2$P$_2$S$_6$ (dark squares). b -- temperature dependence of ultrasound (dark squares) and hypersound (open circles) velocities for transverse TA(ZX) acoustic waves in Sn$_2$P$_2$(Se$_{0.28}$S$_{0.72}$)$_6$ crystals near the LP.\label{fig7}}
\end{figure}

It is known that acoustic behavior near phase transitions is determined by fluctuational effects at $T\!>\!T_0$. Here some contribution from the crystal lattice defects could also appear. In addition at $T\!<\!T_0$, the relaxational interaction between spontaneous polarization and acoustic waves appears. On calorimetric data\cite{bib22} and ultrasound investigations\cite{bib21} for the SPS crystals near the second order phase transition at $T_0\!\approx\!337$~K, the fluctuational contribution in paraelectric phase appears as very small logarithmic corrections. In the ferroelectric phase of SPS crystals, the sound velocity anomaly was successfully described in L--K approximation with single relaxation time for spontaneous polarization.\cite{bib21,bib25_1,*bib25_2} The sound attenuation temperature and frequency dependencies were described\cite{bib25_1,*bib25_2} in the L--K model with accounting of Grunaisen coefficients for low frequency optic and for acoustic modes. Generally, the phase transition in SPS crystals could be considered as example of almost mean field behavior for their acoustic properties. The anomalies of sound velocities for SPS crystal and for the mixed Sn$_2$P$_2$(Se$_{0.28}$S$_{0.72}$)$_6$ crystals with LP composition are compared at Fig.\ref{fig7}. For convenience of comparison of the ultrasound and hypersound velocity temperature behavior, the experimental data are presented in relative coordinates.

It is seen that addition contributions to the $V(T)$ and $\alpha(T)$ anomalies appear near the LP. For LA phonons, such addition contribution was not observed by Brillouin scattering but clearly was found for the ultrasound waves. Here the significant anomalies for the velocity and the attenuation (Fig.\ref{fig4}) appear in the paraelectric phase of $x\!=\!0.28$ crystals. For SPS crystals, the ultrasound attenuation is near 1~cm$^{-1}$ in the paraelectric phase and almost constant at variation of temperature.\cite{bib27}

For TA(ZX) phonons, clearly elastic softening is observed near the LP on Brillouin scattering data. Here the hypersound velocity decreases from 2000~m/s in the paraelectric phase till 900~m/s near phase transition temperature (Fig.\ref{fig3}). The ultrasound data also reveal acoustic softening for the transverse waves that propagate in [001] direction and polarized in [100] direction. For such waves, the velocity decreases from 2100~m/s in the paraelectric phase to 800~m/s at phase transition (Fig.\ref{fig5}). But for the trasverse TA(ZY) waves at the phase transition, the small jump on the velocity temperature dependence is only observed (Fig.\ref{fig6}). It is important to note that the attenuation for TA(ZX) phonons have not relatively strong anomaly near the LP on Brillouin scattering data (Fig.\ref{fig3}), but $\alpha(T)$ dependency for the TA(ZX) phonons strongly rises near the LP as seen from ultrasound data (Fig.\ref{fig5}).

So, really acoustic softening is found for LA(ZZ) and especially for TA(ZX) phonons in the paraelectric phase near the LP in Sn$_2$P$_2$(Se$_{0.28}$S$_{0.72}$)$_6$ crystals. For LA phonons, this softening together with rise of attenuation appears only in ultrasound range. For TA phonons, the elastic softening is observed in hypersound region without strong anomaly of attenuation and in ultrasound region with relatively big anomaly of the acoustic attenuation.

In the ferroelectric phase both for SPS and $x$=0.28 crystals on Brillouin scattering data, the hypersound velocity of LA(ZZ) phonons demonstrate similar behavior which is characteristic for the L--K anomaly. Here the ultrasound velocity temperature dependence almost coincides with hypersound $V(T)$ dependence at low temperatures.

\section{Analysis of experimental data}

The IC phase and LP presence at the $T-x$ diagram of Sn$_2$P$_2$(Se$_x$S$_{1-x}$)$_6$ mixed crystals is related to linear interaction of soft optic and acoustic branches.\cite{bib1_1,*bib1_2} An increase of the order parameter fluctuations is also expected near the LP.\cite{bib7} Both mentioned origins to the sound velocity and attenuation anomalies near the LP in $x\!=\!0.28$ crystals will be analyzed. For the ferroelectric phase, the $V(T)$ and $\alpha(T)$ dependencies are analyzed within L--K model.\cite{bib24}

Let's start with consideration of possible appearance of the linear interaction of soft optic and acoustic phonons in the paraelectric phase at cooling to the LP. The soft optic branch is fully symmetrical ($A'$) in $q_z$ direction of Brillouin zone for the P2$_1$/c space group. For such wave vectors, the acoustic branches for LA(ZZ) and TA(ZX) phonons also found similar symmetry. The linear interaction of soft TO phonons with acoustic phonon branches was clearly observed by inelastic neutron scattering in paraelectric phase of Sn$_2$P$_2$Se$_6$ crystals.\cite{bib10} The strength of this interaction $d$ is in square dependence on the wave vector modulus $q$ ($d\!=\!\mu q^2$). It determine the lattice instability near 0.1$q_{max}$ and the IC modulation appearance at the second order phase transition with $T_i\!\approx\!221$~K in Sn$_2$P$_2$Se$_6$ crystals.\cite{bib2_1,*bib2_2,bib10} In the mean field approximation for the incommensurate phases of type II in proper ferroelectrics, the coupling between the strain $u$ and the order parameter $P$ (polarization) can induce the incommensurate instability, as firstly discussed for the IC phase in quartz.\cite{bib14} The thermodynamic potential in such a case can be written as
\begin{eqnarray}
  \Phi &=& \Phi_0+\frac{A}{2}P^2+\frac{B}{4}P^4+\frac{C}{6}P^6
  +\delta\left(\frac{\partial P}{\partial z}\right)^2 \nonumber\\
  &+&\frac{g}{2}\left(\frac{\partial^2 P}{\partial z^2}\right)^2+\frac12 cu^2+\mu\left(\frac{\partial P}{\partial z}\right)u_{xz}.\label{eq3}
\end{eqnarray}

Here the interaction between the strain and the order parameter is represented by the last term. Such interaction over a wide $q$--range can be described by\cite{bib10}
\begin{equation}\label{eq4}
    d(q)=\frac{\mu q}{1.5\pi}\sin(1.5\pi q)
\end{equation}
and for small $q$ we find $d(q)\!=\!\mu q^2$. The value of $\mu$ was adjusted to fit the parameter $d$ through the Brillouin zone. The effects of discrete lattice were accounted by taking for the acoustic dispersion
\begin{equation}\label{eq5}
    \omega_A=\frac{V_A}{\pi}\sin(\pi q).
\end{equation}

The position of the incommensurate instability is equal to
\begin{equation}\label{eq6}
    q_i^2=-\left(\frac{\delta-\mu^2/V_A^2}{g}\right)\equiv-\frac{\delta^*}{g}.
\end{equation}
In this model, it was put all of the temperature dependence in the parameter $A\!=\!\alpha_T(T\!-\!T_0)$, i.e. all of the soft-mode behavior in the "optic" ($P\!\equiv\!P_x$) fluctuations, while keeping other parameters of potential (\ref{eq3}) fixed in the simulation of the phonons branches at different temperatures.
	
In the mixed Sn$_2$P$_2$(Se$_x$S$_{1-x}$)$_6$ crystals, $\delta^*\!\sim\!(x\!-\!x_{LP})$ and the temperature width of IC phase continuously decreases at $x\!\rightarrow\!x_{LP}$ which is observed experimentally on the $T-x$ diagram.\cite{bib4_1,*bib4_2} As it is expected, the modulation wave vector continuously moves to zero with concentration $x$ lowering to $x_{LP}$. The $q_i^2\!\sim\!(x\!-\!x_{LP})$ dependence was observed by X--ray diffraction\cite{bib2_1,*bib2_2} at selenium content decreasing from 1 till 0.6. For smaller concentrations of selenium, in the interval $0.28\!<\!x\!<\!0.6$, the satellites related to the modulation wave are too close to the main Bragg peaks and those satellites were not resolved at diffraction experiments.\cite{bib2_1,*bib2_2}

From relation (\ref{eq6}), it follows that $q_i\!\rightarrow\!0$ at rise of soft optic branch dispersion coefficient $\delta$, at increasing of elastic modulus $c_{ij}$, or at lowering of interaction constant $\mu$. It was earlier estimated\cite{bib10} that for the Sn$_2$P$_2$(Se$_x$S$_{1-x}$)$_6$ mixed crystals $q_i\!\rightarrow\!0$ at $x\!\rightarrow\!x_{LP}$ if $\delta$ value have been changed from 1.9~THz$^2$($c^*$)$^{-2}$ (in units introduced in Ref.~[\onlinecite{bib10}] where $c^*\!=\!2\pi\times0.1469\AA^{-1}$) at $x\!=\!1$ to 2.42~THz$^2$($c^*$)$^{-2}$ at $x\!=\!0.28$. In this case, the values $\mu\!=\!4.2$~THz$^2$($c^*$)$^{-1}$, $g\!=\!44$~THz$^2$($c^*$)$^{-4}$, $\alpha_T\!=\!0.0018$~THz$^2$\,K$^{-1}$, $V_A\!=\!2.7$~THz($c^*$)$^{-1}$ were constant.

Now we will use these values of $\delta$, $\mu$ and $V_A$ for the analysis of linearly interacting soft optic branch and acoustic branches temperature behavior in the paraelectric phase of the $x\!=\!0.28$ crystal. As was shown earlier,\cite{bib14,bib17} the linear interaction between optic and acoustic phonon branches could be described by the following relations for the frequencies and damping's of soft optic and acoustic phonons (in the case of $\gamma^2\!\ll\!\omega_0^2\!+\!\omega_A^2$):
\begin{eqnarray}
    \omega_{\pm}^2&=&\frac12\left\{\left(\omega_0^2+\omega_A^2\right)\pm\left[\left(\omega_0^2-\omega_A^2\right)^2+4d^2\right]^{1/2}\right\},\label{eq7}\\
   \Gamma_{\pm}&=&\gamma\frac{\omega_A^2-\omega^2_{\pm}}{\omega^2_{\mp}-\omega^2_{\pm}},\label{eq9}
\end{eqnarray}
where $\gamma$ is damping constant for the "bare" optic phonons. Here the dispersion of such optic phonons is
$\omega_0^2\!=\!A\!+\!\delta q^2\!+\!\frac g2q^4$. The dispersion branches for acoustic phonons and interaction parameter $d$ were presented by relations (\ref{eq4}) and (\ref{eq5}). We suppose that only real interaction constant could be accounted.

\begin{figure*}
 \includegraphics*[width=3.4in]{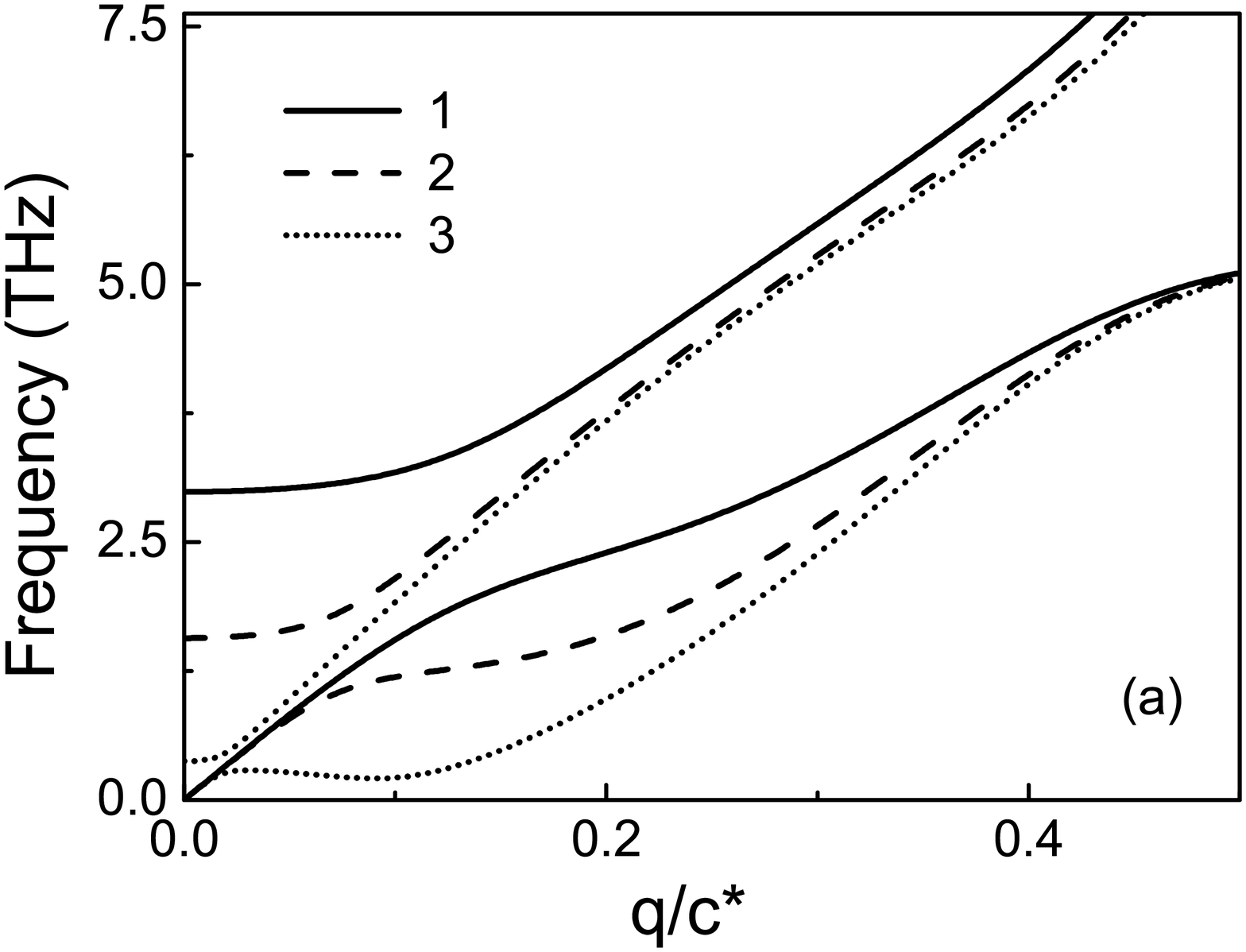}
 \includegraphics*[width=3.4in]{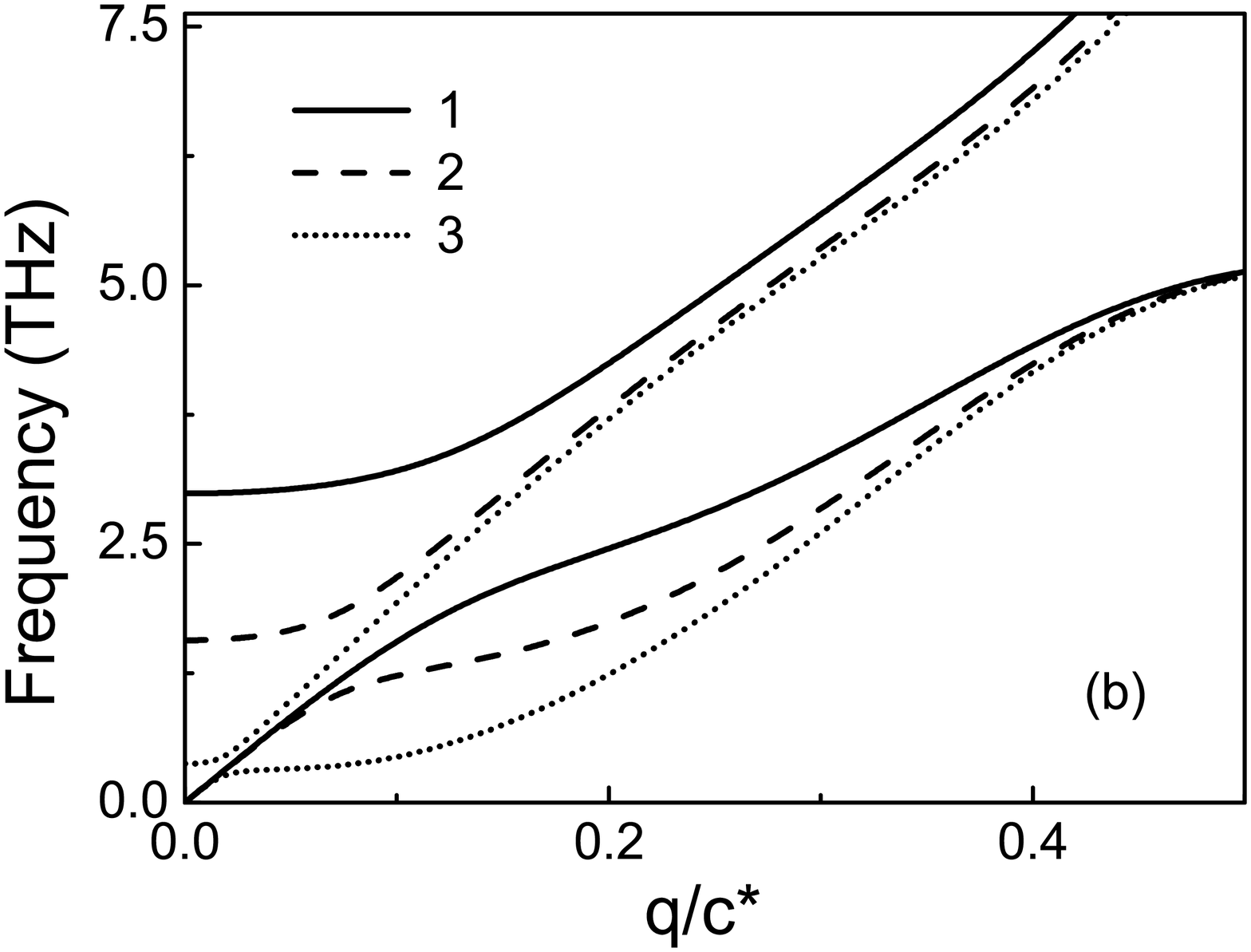}
 \includegraphics*[width=3.4in]{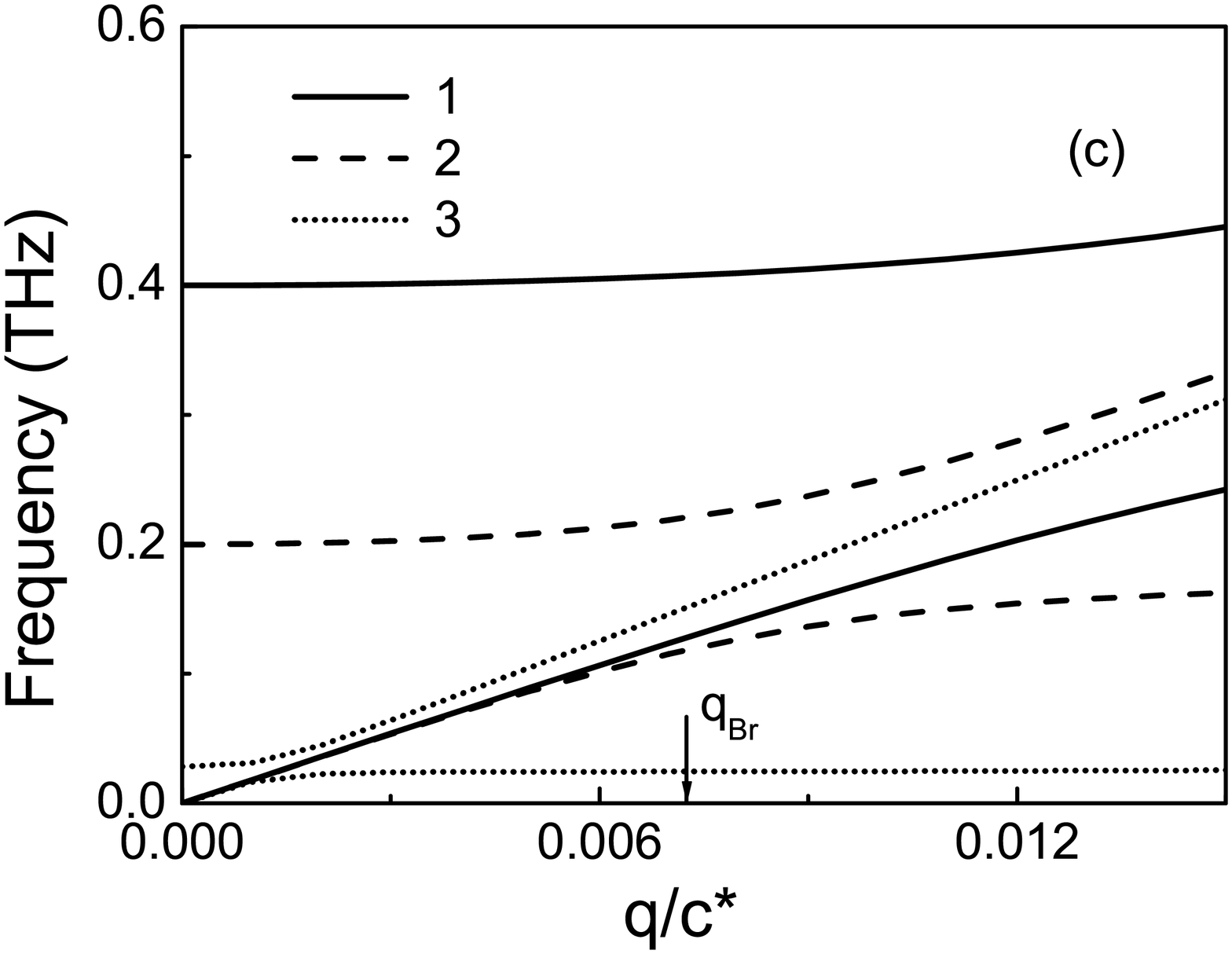}
 \includegraphics*[width=3.4in]{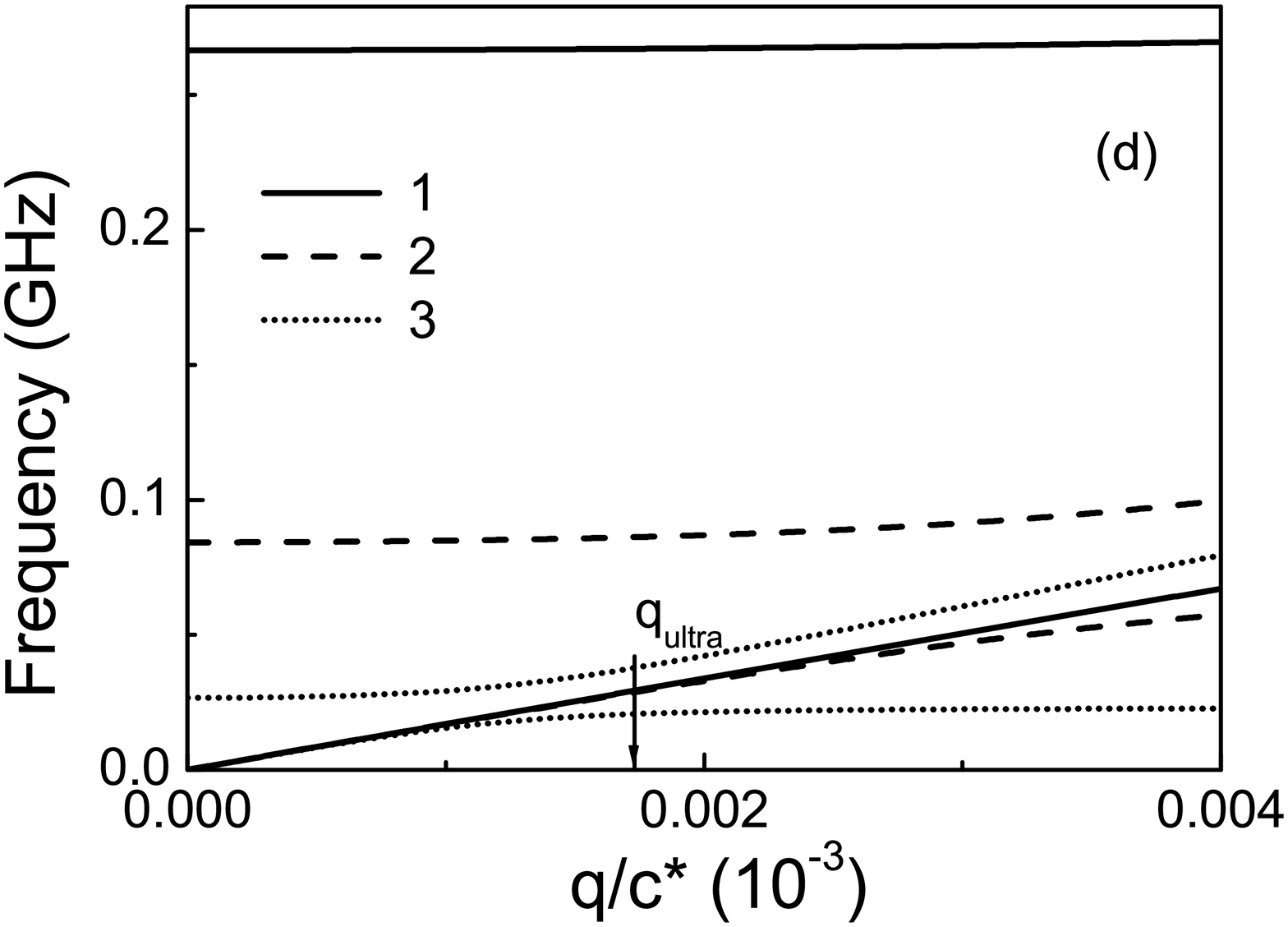}
 \caption{Calculated by relation (\ref{eq7}) temperature evolution of linearly interacting soft optic and acoustic branches across Brillouin zone for the paraelectric phase of Sn$_2$P$_2$Se$_6$ crystal (a) and Sn$_2$P$_2$(Se$_{0.28}$S$_{0.72}$)$_6$ crystal (b) (1 -- $\Delta T\!=\!127$~K, 2 -- $\Delta T\!=\!35$~K, 3 -- $\Delta T\!=\!2$~K). The phonon branches at different temperatures near Brillouin zone center for Sn$_2$P$_2$(Se$_{0.28}$S$_{0.72}$)$_6$ crystal in hypersound (c) (1 -- $\Delta T\!=\!2$~K, 2 -- $\Delta T\!=\!0.5$~K, 3 -- $\Delta T\!=\!0.01$~K) and ultrasound (d) (1 -- $\Delta T$=10$^{-6}$~K, 2 -- $\Delta T\!=\!10^{-7}$~K, 3 -- $\Delta T\!=\!10^{-8}$~K) ranges of frequencies.\label{fig8}}
\end{figure*}

The calculated dispersion curves for the interacting soft optic and acoustic phonons at different temperatures with parameters which have been early determined at fitting of the neutron scattering spectra for the Sn$_2$P$_2$Se$_6$ crystals are represented in Fig.\ref{fig8}(a).\cite{bib10} Here the lattice instability near $q_i\!=\!0.1q_{max}$ is clearly seen that is related to the phase transition into IC phase at $T_i\!\approx\!221$~K. At increase of dispersion parameter $\delta$ to value about 2.42~THz$^2$($c^*$)$^{-2}$, the lattice instability wave vector $q_i$ shifts to the Brillouin zone center that is related to the transition across LP (Fig.\ref{fig8}(b)).

The anticrossing phenomenon is clearly observed at high temperatures where "bare" optic $\omega_o(q)$ and acoustic $\omega_a(q)$ branches cross at enough high values of wave vector $q_{cr}$. The wave vector $q_{cr}$ decreases at $T$ decrease to $T_0$ and interacting constant squarely lowers as $d\!=\!\mu q^2$. By this matter the linear interaction between optic and acoustic branches is clearly observed in wide temperature interval by neutron scattering at $q\!\approx\!0.1q_{max}\!\sim\!10^7$~cm$^{-1}$ for different crystals.\cite{bib11,bib12,bib13} By Brillouin scattering at the wave vector value $q\!\approx\!10^5$~cm$^{-1}$ (or $q\!\approx\!10^{-3}q_{max}$), the linear interaction effect between soft optic and acoustic phonons could be also observed in small temperature interval of paraelectric phase near structural transition (Fig.\ref{fig8}(c)). For ultrasound waves with $q\!\approx\!10^2$~cm$^{-1}$, such interaction effects are expected to be very small (Fig.\ref{fig8}(d)).

\begin{figure}
 \includegraphics*[width=3.4in]{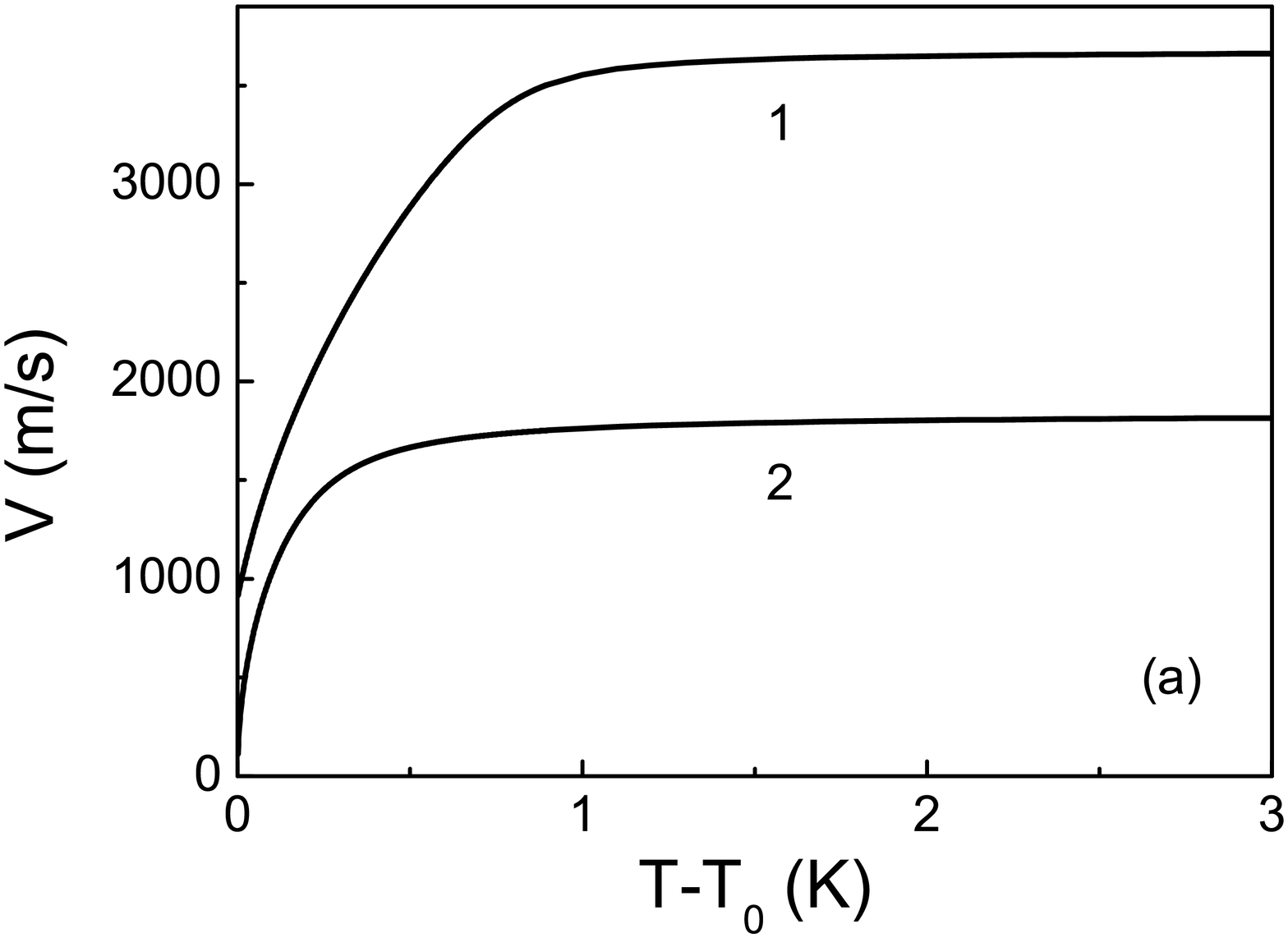}
 \includegraphics*[width=3.4in]{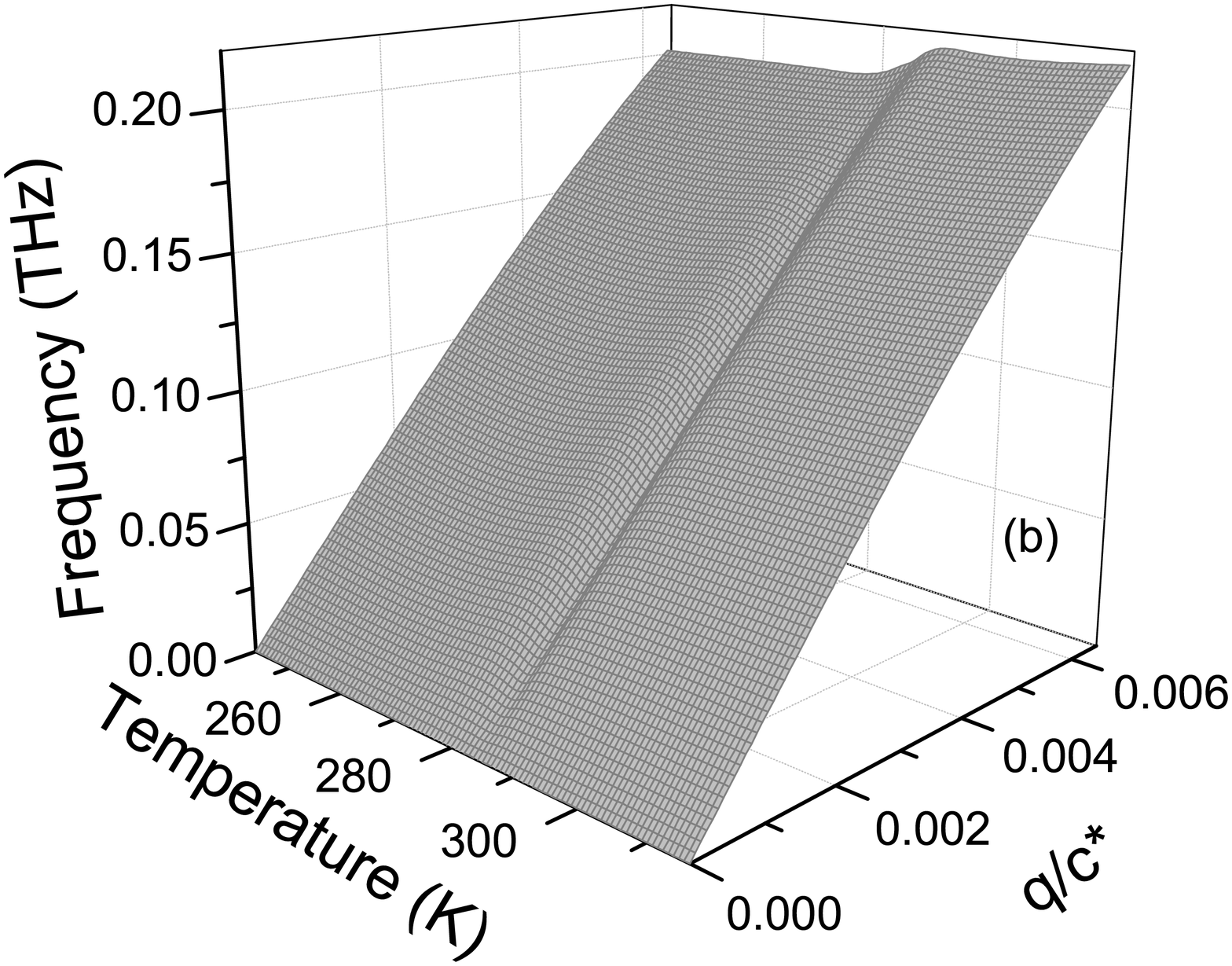}%
 \caption{a --- determined from relation (\ref{eq7}) temperature dependencies of hypersound velocity for the longitudinal LA(ZZ) with $V_{LA}\!=\!2V_{TA}$ (1) and transverse TA(ZX) (2) acoustic phonons in paraelectric phase of Sn$_2$P$_2$(Se$_{0.28}$S$_{0.72}$)$_6$ crystal; b --- calculated temperature evolution of longitudinal acoustic branch near the LP in  Sn$_2$P$_2$(Se$_{0.28}$S$_{0.72}$)$_6$ crystal.\label{fig9}}
\end{figure}

Indeed, as is shown at Fig.\ref{fig9}(a), the calculated temperature dependence for the phase hypersound velocity demonstrates elastic softening at cooling till several degrees to $T_0$. But for the ultrasound phase velocity, such acoustic softening could be expected in very small temperature interval near $T_0$, that practically could not be observed. General view of the acoustic branch temperature transformation at transition across the LP (Fig.\ref{fig9}(b)) demonstrates changing of curvature sign for the acoustic velocity dispersion --- far from the LP  $V_{ultra}\!>\!V_{hyper}$ and near the LP in contrary $V_{ultra}\!<\!V_{hyper}$. Such reverse in the acoustic dispersion  obviously is clear peculiarity of the lattice dynamics near the LP.

\begin{figure}
 \includegraphics*[width=3.4in]{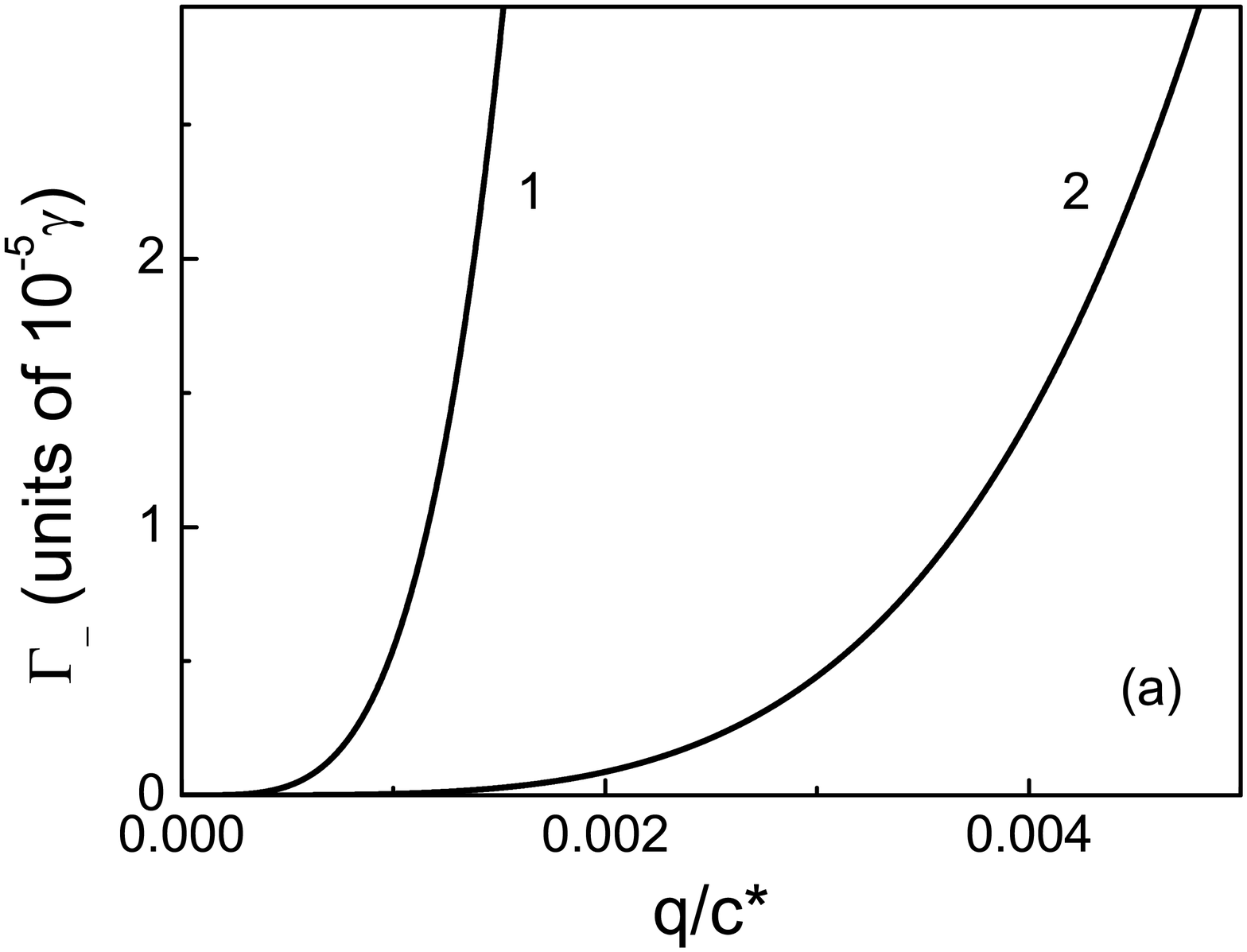}
 \includegraphics*[width=3.4in]{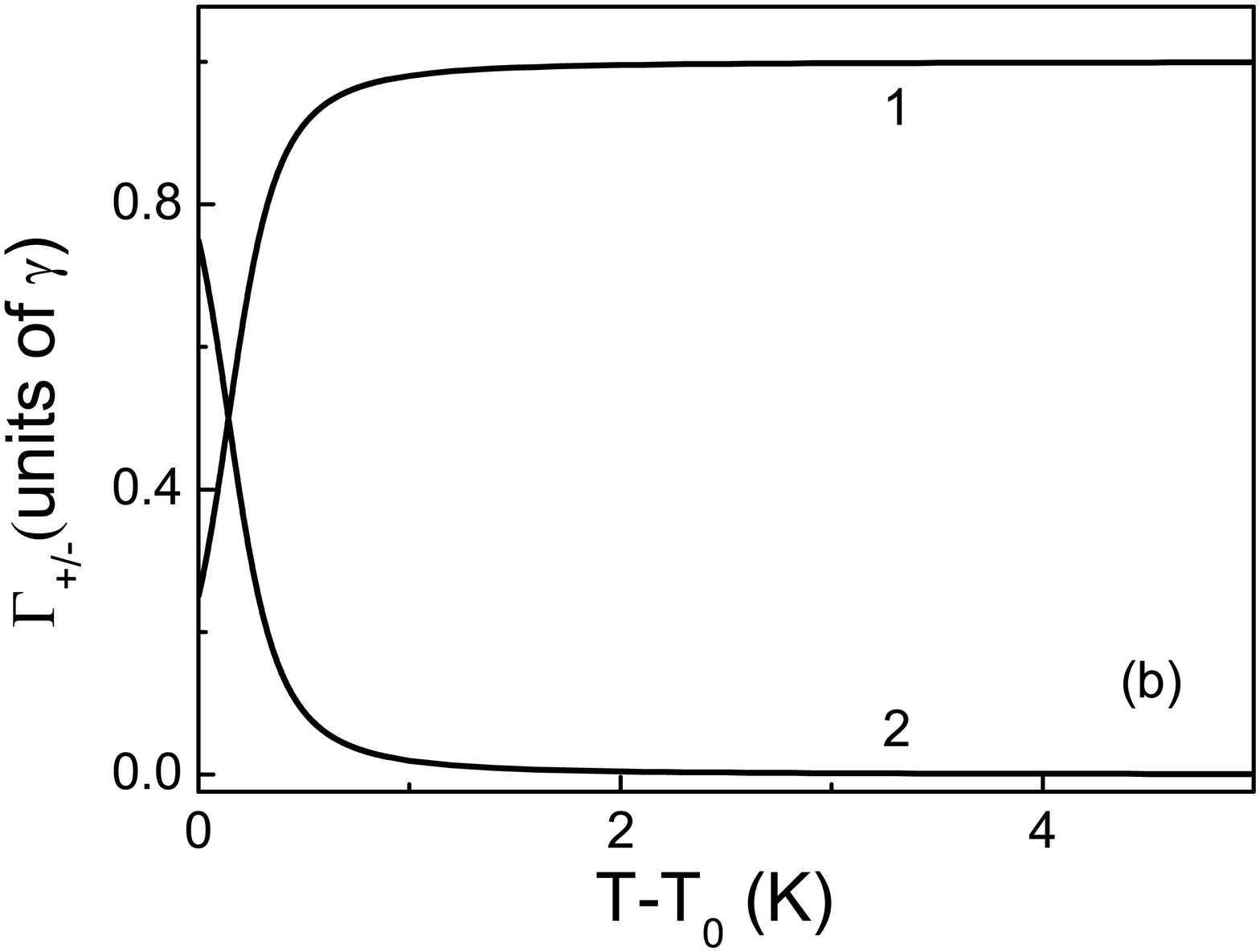}%
 \caption{Calculated by relation (\ref{eq9}) for paraelectric phase of Sn$_2$P$_2$(Se$_{0.28}$S$_{0.72}$)$_6$ crystal wave number dependencies (a) of acoustic phonons damping at two temperatures $T_{LP}\!+\!1$~K (1), $T_{LP}\!+\!10$~K (2), and  temperature dependencies (b) of acoustic (1) and optic (2) phonons damping at wave number $q_{Br}\!\approx\!10^5$~cm$^{-1}$.\label{fig10}}
\end{figure}

Linear interaction of optic and acoustic phonon branches determines strong wave vector dependence of its damping constant that was calculated by relations (\ref{eq9}) for different temperature distance to the LP (Fig.\ref{fig10}). For the acoustic phonons, the damping constant is square function of its wave vector amplitude with growing coefficient at approaching to $T_{LP}$. From calculations, it follows a redistribution of the damping value from optic phonons to acoustic one at temperature variation for investigated by Brillouin scattering wave vector.

\begin{figure}
 \includegraphics*[width=3.4in]{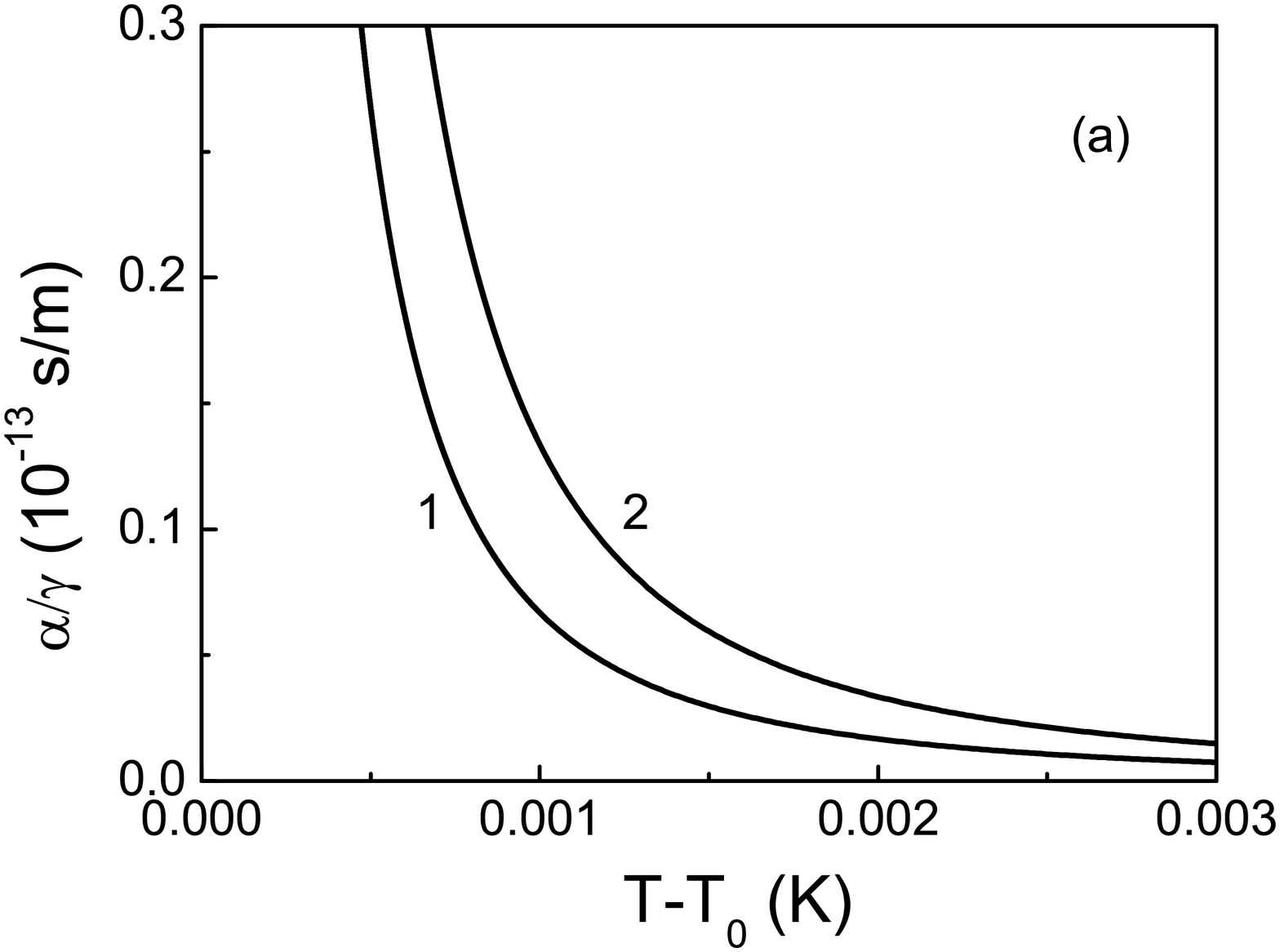}
 \includegraphics*[width=3.4in]{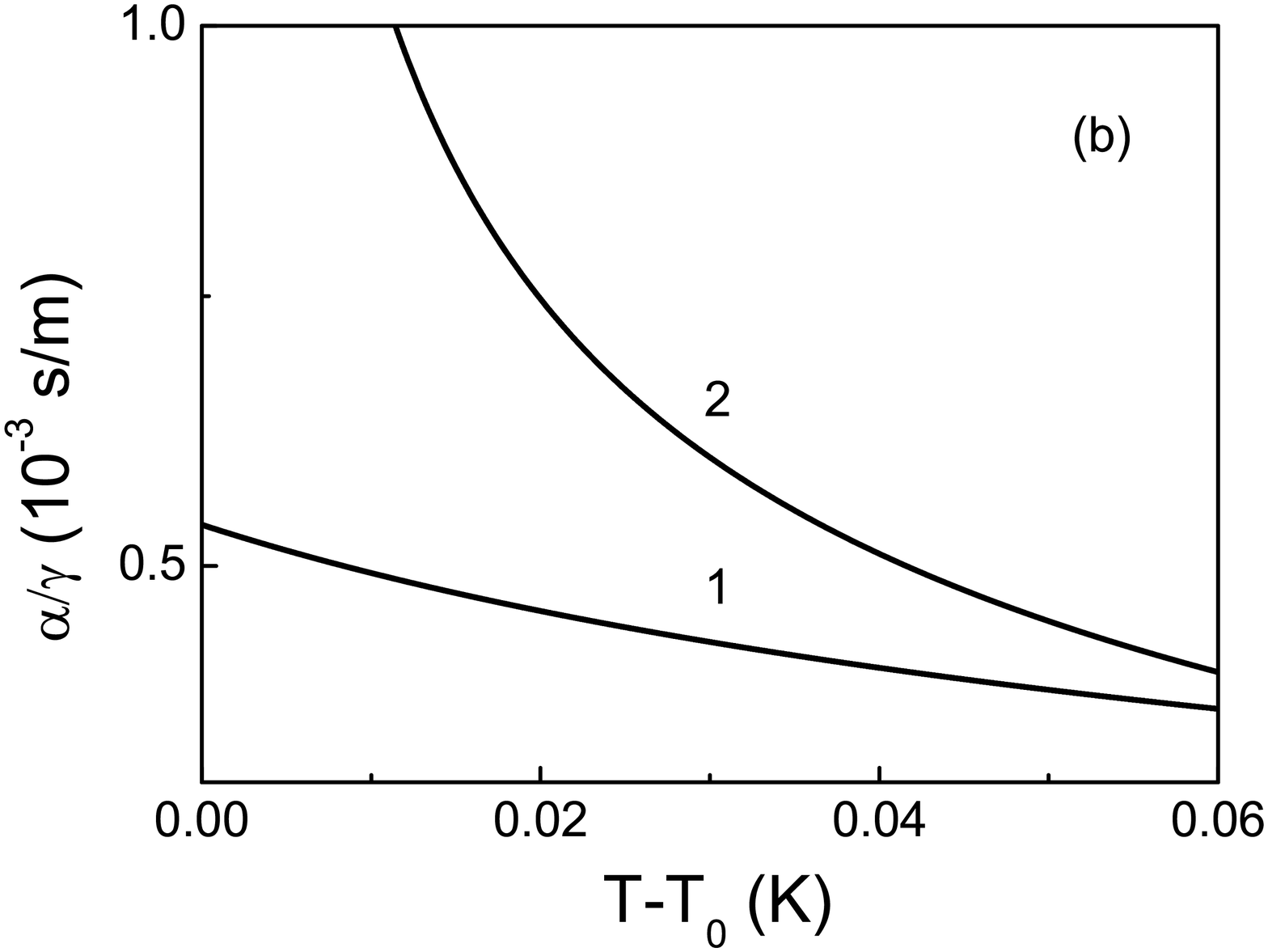}%
 \caption{Calculated using relation (\ref{eq9}) temperature dependencies of ultrasound (a) and hypersound (b) attenuation for longitudinal (1) and transverse (2) phonons in paraelectric phase of Sn$_2$P$_2$(Se$_{0.28}$S$_{0.72}$)$_6$ crystal.\label{fig11}}
\end{figure}

The temperature intervals for anomalous rise of sound attenuation, that are induced by linear coupling with damped soft optic phonons, are expected to be in several degrees interval for hypersound waves and in mKelvins region near $T_0$ for ultrasound waves (Fig.\ref{fig11}), similarly to the temperature intervals of anomalous behavior of its velocities.

As it was found experimentally for the $x\!=\!0.28$ crystals, the ultrasound velocity $V(T)$ anomalously softens and attenuation coefficient $\alpha(T)$ increases in wide enough temperature range of order 10~K above $T_0$ in the paraelectric phase (Figs.\ref{fig4} and \ref{fig5}). It is interesting to note that transverse TA(ZX) acoustic waves demonstrate similar softening at cooling in paraelectric phase in both Brillouin scattering and ultrasound experiments (Fig.\ref{fig7}). Probably, so clearly observed acoustic softening near the LP is determined by increasing of the order parameter fluctuations in its vicinity.

The fluctuational anomalies for sound velocity and attenuation near the ferroelectric phase transitions theoretically were analyzed in several papers.\cite{bib18,bib19,bib20} It was found that these anomalies could be in correlation with temperature dependence of heat capacity $c(T)$. It was proposed possibility of the linear relation between sound velocity and heat capacity anomalies ($V(T)\!\sim\!c(T)$) and square relation between sound attenuation and heat capacity temperature dependencies ($\alpha(T)\!\sim\!c(T)^2$). Such relations were experimentally confirmed for the NaNO$_2$ and TGS ferroelectric crystals.\cite{bib31_1,*bib31_2,bib32}

The heat capacity of SPS crystals was investigated in several papers.\cite{bib33,bib34} On heat diffusivity investigations\cite{bib22}, the heat capacity in the paraelectric phase of SPS crystals demonstrates logarithmic corrections similarly to the temperature dependence of longitudinal ultrasound velocity.\cite{bib21} So, for SPS crystals in the paraelectric phase near $T_0$ obviously the proportionality $V(T)\!\sim\!c(T)$ is observed.

\begin{figure}
 \includegraphics*[width=3.4in]{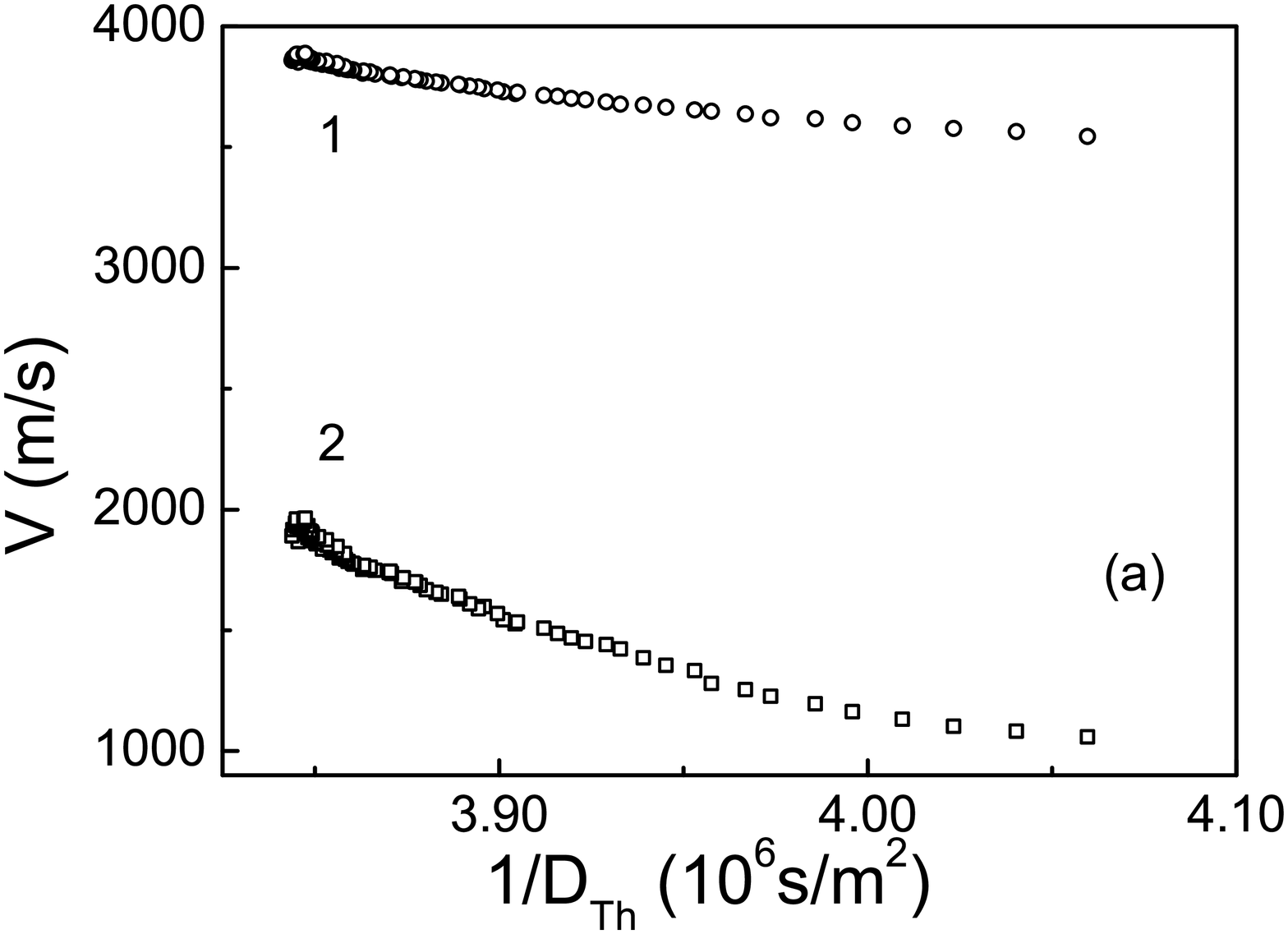}
 \includegraphics*[width=3.4in]{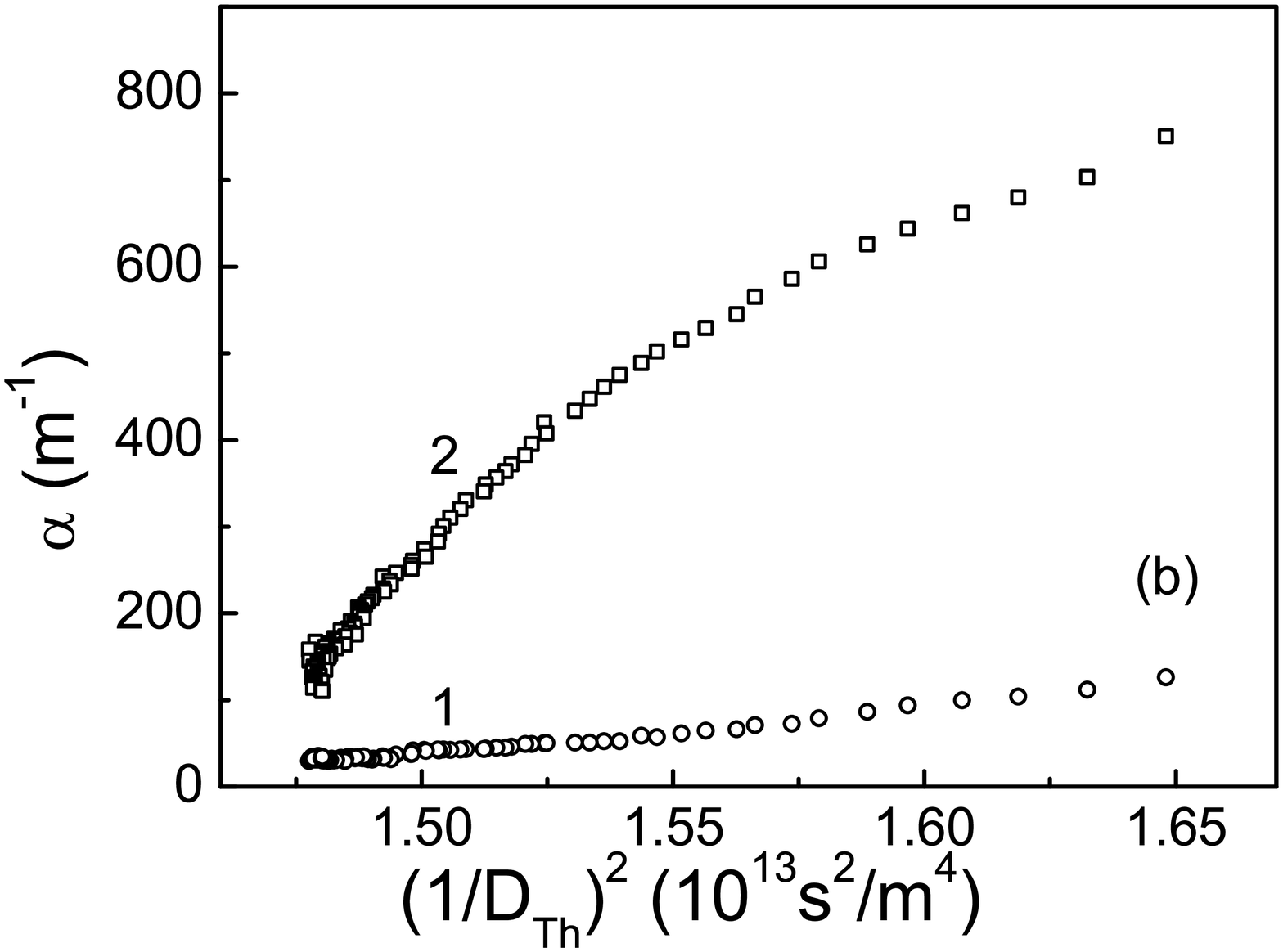}%
 \caption{Relations between ultrasound velocity and heat capacity $V(T)\!\sim\!c(T)\!\sim\!\frac{1}{D_{Th}(T)}$ (a) and between  ultrasound attenuation and heat capacity $\alpha(T)\!\sim\!c(T)^2\!\sim\!\frac{1}{D_{Th}(T)^2}$ (b) for longitudinal (1) and transverse (2) acoustic waves in temperature interval 1~K$\!<\!T\!-\!T_{LP}\!<\!10$~K of paraelectric phase in Sn$_2$P$_2$(Se$_{0.28}$S$_{0.72}$)$_6$ crystal.\label{fig12}}
\end{figure}

We have used the heat diffusivity data\cite{bib35} for analysis of relation between temperature anomalies of heat capacity and ultrasound velocity and attenuation coefficient in the paraelectric phase of $x\!=\!0.28$ crystals. Here it was accounted that the heat capacity $c$ is proportional to the reciprocal of heat diffusivity coefficient $D_{th}$. It is shown in Fig.\ref{fig12} that the temperature dependence of ultrasound velocity for the longitudinal and transverse waves is proportional to the heat capacity temperature variation in enough wide interval 1~K$\!<\!T\!-\!T_0\!<$10~K. So, relation $V(T)\!\sim\!c(T)$ could be considered as evidence of fluctuational origin of observed acoustic softening near the LP in Sn$_2$P$_2$(Se$_x$S$_{1-x}$)$_6$ ferroelectrics.

For ultrasound attenuation, the relation $\alpha(T)\!\sim\!c(T)^2$ is illustrated on example of longitudinal and transverse acoustic phonons in $x\!=\!0.28$ crystal (Fig.\ref{fig12}(b)). Such relation exists in smaller temperature interval in compare with early discussed relation between sound velocity and heat capacity. But in any case, it could be obviously supposed that observed ultrasound anomalies in the paraelectric phase of $x\!=\!0.28$ crystals near the LP are determined by developed fluctuations of the order parameter. Such a wide fluctuational temperature interval was earlier predicted for the LP in Sn$_2$P$_2$(Se$_x$S$_{1-x}$)$_6$ ferroelectrics.\cite{bib7,bib36}

Here it could be also pointed that for the SPS crystals with three--well potential the lattice instability is related to nonlinear interaction of fully symmetrical $A_g$ and soft optic $B_u$ modes.\cite{bib37} Such interaction of $A_gB_u^2$ type induces some softening of the lowest energy $A_g$ optic modes at cooling to $T_0$. Between $A_g$ optic phonons and acoustic phonons LA(ZZ) and TA(ZX) ,the linear interaction is possible at $q\!\rightarrow\!0$. This coupling between strongly anharmonic optic phonons and acoustic waves is obviously observed by fluctuational anomalies of velocity and attenuation for ultrasound in $x\!=\!0.28$ crystals near the LP.

The fluctuational mechanism of sound anomalies in the lowest order is related to the three--phonon processes of $P^2u$ type. Such phonon scattering processes determine the thermal relaxation of phonon subsystem of solids and it is accounted in Akhieser's theory for sound attenuation.\cite{bib28_1,*bib28_2,bib30} The interaction of $P^2u$ type originates the linear interaction of optic and acoustic phonons at deviation from the Brillouin zone center. The interaction of $\mu(\frac{\partial P}{\partial z})u_{xz}$ type appears at $q_z\!>\!0$, which could be considered as linearization of the nonlinear relation $P^2u$ by the wave vector $q_z$. At transition into ferroelectric phase, the spontaneous polarization appears and its static value induces a linearization of relation between order parameter and deformation in the form $P_0Pu$. Such so called linearized electrostriction is an origin of the L--K mechanism\cite{bib24} of sound velocity and attenuation anomalies which will be considered below.

In the ferroelectric phase of monoclinic crystals for the longitudinal sound along [001] direction in the symmetry plane at L--K approach, the relations for temperature behavior of sound velocity and attenuation could be found from thermodynamic potential (\ref{eq3}) without gradient invariants and at adding of electrostriction related invariants $q_{ijk}P^2u$ and $r_{ijkl}P^2u^2$. In such a case, it follows the expressions for velocity
\begin{eqnarray}
   V_{33}^2&=&V^2_{33\infty}-\frac{1}{1+\omega^2\tau^2}\Biggl[\frac{2q^2_{113}}{\rho B\sqrt{1-\frac{4A C}{B^2}}}\nonumber\\
&+&\frac{r_{1133}B}{2C\rho}\left(\sqrt{1-\frac{4A C}{B^2}}-1\right)\Biggr],\label{eq10}
\end{eqnarray}
and for the relaxational attenuation
\begin{equation}\label{eq11}
    \alpha=\frac{V_{\infty}-V^2}{2V^3}\frac{\omega^2\tau}{1+\omega^2\tau^2}.
\end{equation}
Using these relations, the temperature dependencies of sound velocity and attenuation were calculated for temperature interval of the ferroelectric phase and it is compared with experimental data (Fig.\ref{fig2}). At fitting of the $V(T)$ dependence in the ferroelectric phase of investigated crystals, the earlier determined\cite{bib4_1,*bib4_2} coefficient of the thermodynamic potential (\ref{eq3}) were used: $\alpha_T\!=\!1.6\!\times\!10^6$~J\,m\,C$^{-2}$\,K$^{-1}$, $B\!=\!3.98\!\times\!10^8$~J\,m$^5$\,C$^{-4}$, $C\!=\!4.905\!\times\!10^{10}$~J\,m$^9$\,C$^{-6}$. It was also found the electrostriction coefficients $q_{113}\!=\!3.5\!\times\!10^9$~J\,m\,C$^{-2}$, $r_{1133}\!=\!3.9\!\times\!10^{10}$~N\,m$^2$\,C$^{-2}$ and relaxation time $\tau_0\!=\!7.87\!\times\!10^{-11}$~s\,K for temperature dependence $\tau\!=\!\tau_0/(T_0\!-\!T)$. For the sound velocity, the calculated behaviour coincides with experimental data obtained by Brillouin scattering. But calculated dependence $V(T)$ strongly deviates from ultrasound velocity temperature behavior near the phase transition. Such strong sound velocity dispersion near $T_{LP}$ at crossover from ultrasound to hypersound interval is obviously related to the fluctuational effects near the LP, what was early discussed for the paraelectric phase. Now could be stressed that we found clear demonstration of very strong rise of long--wave fluctuations, observed in ultrasound wave length interval, in comparison with relatively small developments of the short--range fluctuations in the wave length scale of hypersound range.

The calculated by formula (\ref{eq11}) sound attenuation in the ferroelectric phase of $x\!=\!0.28$ crystals near maxima of the $\alpha(T)$ dependence is almost twice larger than observed by Brillouin scattering (Fig.\ref{fig2}). For SPS crystals such disagreement was explained\cite{bib25_1,*bib25_2} by including the mode Gruneisen coefficients as interaction parameters instead of electrostriction coefficients. In such estimation, several low frequency optic modes, which interact with acoustic waves, were taken into account. Its contribution to the sound velocity critical anomaly adequately is described in the ferroelectric phase using electrostriction coupling. But for the sound attenuation temperature anomaly, only the lowest frequency optic modes have the major contribution and Gruneisen coefficients of these modes satisfactory describe optic--acoustic dissipative interaction.\cite{bib28_1,*bib28_2,bib29,bib30}

\section{Conclusions}
For proper uniaxial Sn$_2$P$_2$(Se$_x$S$_{1-x}$)$_6$ ferroelectrics, the temperature dependencies of sound velocity and attenuation in the vicinity of Lifshitz point ($x_{LP}\!\approx\!0.28$ and $T_{LP}\!\approx\!284$~K) for the phonons that propagate along direction of modulation wave vector (in the incommensurate phase at $x\!>\!x_{LP}$) were investigated by Brillouin scattering and by ultrasonic pulse--echo method. The elastic softening was found by Brillouin scattering at cooling in the paraelectric phase to $T_{LP}$, as remarkable lowering of hypersound velocity for the transverse acoustic phonons that are polarized in crystallographic plane containing the vector of spontaneous polarization and the wave vector of modulation. Such softening was observed for both longitudinal and transverse acoustic waves for the ultrasound frequencies. Also, strong rise of attenuation for these waves have been observed. By analysis of phonon spectra with accounting of the thermodynamic properties anomalous behavior near the LP in investigated crystals, it was shown that such elastic softening in hypersound range is induced mostly by linear interaction between soft optic and acoustic phonon branches. For the ultrasound frequency region, the elastic softening and growth of acoustic attenuation are related to developed order parameter fluctuations near the LP. Generally, it was demonstrated that linear interaction for short--range optic and acoustic fluctuations and nonlinear interaction for these long--range fluctuations are dominated at approaching to the LP. In the ferroelectric phase, the Landau--Khalatnikov model explains temperature dependence of hypersound velocity.

\begin{acknowledgments}
This work was supported by the Ukrainian--Lithuanian project "Electronic properties and phase transitions in phosphorous
chalcogenide semiconductors".
\end{acknowledgments}
\bibliographystyle{apsrev4-1}
\bibliography{bibliogr}

\end{document}